\documentclass[journal]{new-aiaa}
\usepackage[utf8]{inputenc}
\usepackage{textcomp}

\usepackage{graphicx}
\usepackage{amsmath}
\usepackage[version=4]{mhchem}
\usepackage{siunitx}
\usepackage{longtable,tabularx}
\usepackage{outlines}
\usepackage{comment}
\usepackage{color}
\usepackage{multirow}
\usepackage{subfigure}
\usepackage{ulem}
\usepackage[textsize=tiny, 
color=blue!30]{todonotes}
\usepackage{lineno}
 
\newcommand{\subrm}[1]{{_{\mathrm{#1} } }} 
\newcommand{\subsuprm}[2]{{ _{\mathrm{#1}}^{\mathrm{#2} } }} 
\newcommand{\einsteincoeff}{\mathcal{A}_{ji}}
\newcommand{\specnot}[5]{ \textrm{#1} \; {}^{#2}{\textrm{#3}}_{#4}^{#5}}

\newcommand{\redcam}{``Red''}
\newcommand{\ircam}{``IR''}

\newcommand{\speedoflight}{\textrm{c}} 
\newcommand{\boltzmannconstant}{\textrm{k}\subrm{B}}

\newcommand{\Trot}{T\subrm{rot}} 
\newcommand{\Tvib}{T\subrm{vib}}

\title{Combined Absorption and Emission Study of Air Species in NASA EAST}

\author{Jesse W. Streicher \footnote[1]{Research Scientist, Department of Mechanical Engineering.}, Efaine Chang\footnote[2]{Graduate Research Assistant, Department of Aeronautics and Astronautics. AIAA Student Member.}, Dylan L. Drescher \footnote[3]{Graduate Research Assistant, Department of Mechanical Engineering.}, Christopher L. Strand \footnote[4]{Senior Research Engineer, Department of Mechanical Engineering.}, Ronald K. Hanson \footnote[5]{Clarence J. and Patricia R. Woodard Professor of Mechanical Engineering, Department of Mechanical Engineering. AIAA Fellow.}}

\affil{Stanford University, Stanford, California 94305, USA}

\author{Andrea Fagnani\footnote[6]{NASA Postdoctoral Program Fellow, Aerothermodynamics Branch}, Augustin C. Tib\`{e}re-Inglesse\footnote[7]{NASA Postdoctoral Program Fellow, Aerothermodynamics Branch}}
\affil{ORAU at NASA Ames Research Center, Moffett Field, CA, 94035, USA}

\author{Brett A. Cruden\footnote[8]{Senior Research Scientist, Aerothermodynamics Branch. AIAA Associate Fellow.}}
\affil{AMA, Inc., at NASA Ames Research Center, Moffett Field, California 94035, USA}

\begin{document}

\maketitle

\begin{abstract}
\vspace{-6mm}
A suite of tunable diode laser absorption spectroscopy (TDLAS) diagnostics targeting rovibrational states of nitric oxide (NO) in the mid-infrared (MIR) and electronic states of atomic oxygen (O) and nitrogen (N) in the near-IR (NIR) was combined with optical emission spectroscopy (OES) for overlapping wavelengths at the NASA Electric Arc Shock Tube (EAST) facility for three selected high-temperature air experiments.
Absorption features were probed at rates up to 1 MHz to provide quantitative, time-resolved measurements of temperature and species number density profiles behind incident shocks, enabling comparisons to spatially-resolved emission measurements, with the two typically agreeing within their respective uncertainties.
Inferred time-histories are further compared to incident shock simulations from two one-dimensional models -- one leveraging the Park two-temperature rates and the other more recent rate values -- as well as the Data Parallel Line Relaxation (DPLR) code.
For the two low-velocity cases (3.81 and 3.98 km/s) in which NO is probed, inferred temperatures are typically well-captured by all three models, although the NO number densities show stronger variation, agreeing better for the model with updated NO formation and dissociation rates.
Finally, one high-velocity case (7.03 km/s) demonstrates the simultaneous resolution of three O-atom and six N-atom electronic states, with all states showing a steady increase in number density over time -- possibly due to shock trajectory effects.
Together, this combined absorption and emission study demonstrates resolution of many states of nonequilibrium air species, while directly comparing ground- and excited-state populations via the combined diagnostic strategy.

\end{abstract}

\section*{Nomenclature}
{\renewcommand\arraystretch{0.9}
\noindent\begin{longtable*}{@{}l @{\quad=\quad} l@{}}
$A_k$	            &	integrated absorbance of transition k, cm$^{-1}$\\ %%
$\mathcal{A}_{ji}$	&	Einstein A coefficient of transition j to i, $\mathrm{s}^{-1}$\\ %%
$\textrm{c} $       &   speed of light, m/s\\ %%
$C_{ji}$            &   emission linestrength coefficient of transition j to i, $\SI{}{\watt/(\steradian)}$\\%%
$E_{\textrm{rot}}$  &   lower-state rotational energy, cm$^{-1}$ or J\\ %%
$E_{\textrm{vib}}$  &   lower-state vibrational energy, cm$^{-1}$ or J\\ %%
$E_{j}$             &   energy of electronic energy level j, eV or J\\ %%
$D$                 &   EAST facility internal diameter (10.16 cm)\\ %%
$f_{gate}$          &   detector gate function \\ %%
$F_{\lambda}$       &   sensor response \\ %%
$g_{j}$, $g_{i}$    &   degeneracy of the ground (i) or excited (j) state of a transition\\ %%
$h$                 &   Planck's constant, Js\\ %%
$I_0$	            &	incident intensity\\ %%
$I_{\textrm{t}}$	&	transmitted intensity\\ %%
$J"$	            &	lower-state rotational quantum number\\ %%
$k_B$               &   Boltzmann constant, J/K\\ %%
$K_{ji}$            &   absorption linestrength coefficient of transition j to i, $\SI{}{\centi \meter \cubed}$ \\%%
$l_{abs}$	        &	absorbing path length, cm\\ %%
$L_{\lambda_p}$	    &	spectral radiance, $\SI{}{\watt / (\centi \meter\squared \steradian \hspace{1mm} \nano\meter)}$\\ %%
$L_{BB}(T)$	    &	blackbody spectral radiance, $\SI{}{\watt / (\centi \meter\squared \steradian \hspace{1mm} \nano\meter)}$\\ %%
$M$	                &	molar mass of absorbing species, amu\\ %%
$n_{j}$, $n_{i}$     &   number density of the ground (i) or excited (j) state of a transition, mol cm$^{-3}$\\ %%
$n_{total}^s$       &   total number density of species s, mol cm$^{-3}$\\ %%
$P$	                &	pressure, atm or Torr\\ %%
$Q$                 &   partition function\\ %%
$r_{gate}$          &   gating reciprocity factor \\ %%
$R_{\lambda_p}$     &   absolute spectral radiance responsivity \\ %%
$S_k$ 	            &	temperature-dependent linestrength of transition k, $\mathrm{cm^{-2}atm^{-1}}$\\ %%
$T$                 &   temperature, K\\ %%
$T_{\mathrm{elec}}$                & electronic temperature, K\\ %%
$T_{\mathrm{LTE}}$                 & local thermal equilibrium temperature, K\\ %%
$T_{\mathrm{rot}}$                 & rotational temperature, K\\ %%
$T_{\mathrm{trans}}$               & translational temperature, K\\ %%
$T_{\mathrm{vib}}$                 & vibrational temperature, K\\ %%
$t$                 &   post-shock time, s\\%%
$U_{p}$                 & digitalized intensity counts on pixel, p\\ %%
$v"$	            &	lower-state vibrational quantum number\\ %%
$v'$	            &	upper-state vibrational quantum number\\ %%
$V_{shock}$	            &	shock velocity, km/s\\ %%
$w$	                &	uncertainty weight parameter\\ %%
$X_\mathrm{abs}$	    & mole fraction of the absorbing species\\ %%
$x$	                    & post-shock distance, cm\\ %%
$\alpha_{\nu,\lambda}$  & absorbance at frequency $\nu$ or wavelength $\lambda$\\ %%
$\alpha_{det}$          & detector constant \\ %%
$\Delta E_{ji}$         & energy difference between states j and i, cm$^{-1}$, eV, J\\ %%
$\Delta\nu_C$           & collisional full-width at half-maximum, cm$^{-1}$\\ %%
$\Delta\nu_D$           & Doppler full-width at half-maximum, cm$^{-1}$\\ %%
$\varepsilon\subrm{\lambda}$        & spontaneous emission intensity, $\SI{}{\watt /(\centi \meter\cubed \steradian \hspace{1mm} \nano\meter)}$\\ %%
$\Theta$                & instrument throughput \\ %%
$\kappa_{\lambda}$      &  absorption coefficient, $\SI{}{\per \centi \meter}$\\ %%
$\lambda_{ji}$          & center wavelength defining the transition between states j and i  , nm\\ %%
$\phi_{ij}$, $\phi_{k}$	  & lineshape function of transition k between states i and j, cm for $\phi_{k}(\nu)$, cm$^{-1}$ for $\phi_{i,j}(\lambda)$\\ %%
$\nu_{0,k}$	            & vacuum line center of transition k, cm$^{-1}$\\ %%
$\chi_{\ce{BB}}$	            & background continuum scaling factor\\ %%
\end{longtable*}}

%\linenumbers

\graphicspath{{figures/}}
\section{Introduction}
Hypersonic vehicles, whether aircraft traveling through Earth's atmosphere or spacecraft reentering Earth's atmosphere, experience intense heat loads from the flow following a strong bow shock on the forebody. The corresponding thermal protection system (TPS) for such vehicles is often over-designed to manage this heating with a sufficient margin of safety, due to the uncertainties associated with the complex molecular physics and chemistry occurring within this region \cite{physics_1, physics_2, physics_3}. While these phenomena can be simulated, model validation requires experimental, gas-phase measurements to supplement surface measurements in relevant flows \cite{measurements_1, measurements_2, measurements_3}. Therefore, a wide range of hypersonic ground test facilities and high-fidelity optical diagnostics are utilized for validating air chemistry models and predicting the interactions of the shock heated gas with realistic TPS materials.
\par Of such ground test facilities, the NASA Ames Electric Arc Shock Tube (EAST) is one which can replicate relevant velocities and free stream densities, based on density equivalent pressure. This in turn allows for measurement of gas composition and reaction kinetics that have similarity for shock layer radiation. Although EAST can access a wide range of conditions, the data collected in Test Series 66 \cite{cruden2024b} cover a regime of lower shock velocities to access conditions relevant to hypersonic aircraft. The EAST facility traditionally infers absolute radiance through optical emission spectroscopy (OES) \cite{oes}. However, the interpretation of gas properties (other than radiation magnitude) from OES data relies on excitation rates from modeling, which may carry significant uncertainty \cite{assumptions}. Furthermore, emission signals for low-speed shock experiments are much weaker than for high-speed shocks. This motivates using tunable diode laser absorption spectroscopy (TDLAS) in addition to OES, to directly track molecular temperatures and species concentrations behind the shock front by probing lower states, which retain strong signals in low-speed shock experiments. The data obtained using TDLAS techniques may then be used in combination with OES to build, validate, and refine models for kinetics and radiative emission, as the two techniques provide complementary data probing upper- and lower-states.
\par This work utilizes TDLAS in the mid-infrared (MIR) and near-infrared (NIR) to provide high signal-to-noise ratio (SNR), temporally-resolved measurements of temperature and state-specific populations to complement EAST OES diagnostics and inform radiation models. Over the whole of Test Series 66, conditions encompass incident shock velocities 3.0-7.1 km/s and fill pressures 1.4-10.0 Torr in air, including high pressures at which chemical equilibrium is captured, and low pressures where chemical nonequilibrium is observed. The full test matrix was additionally divided into two subsets, where at velocities less than 5 km/s, the TDLAS diagnostic probes nitric oxide (NO); at higher velocities electronically excited atomic oxygen and nitrogen (O, N) are probed. Within these two regimes, example results are provided for two low-speed experiments (3.98 and 3.81 km/s) with varying fill pressure (9.96 and 2.25 Torr, respectively) that probe NO under equilibrium and nonequilibrium condition, and one high-speed experiment (7.03 km/s) is provided for the investigation of O and N electronic states under equilibrium conditions.

%=======================================================================================================
\section{Experimental Methods}
In this work, the EAST facility provides controlled, repeatable test conditions, the IR-based absorption diagnostics provide high time resolution and high SNR measurements for lower states of absorption transitions, and the emission diagnostics probe the corresponding upper-state populations, for a given wavelength of an optical transition. In some cases, TDLAS and OES can probe the same state populations, albeit at different wavelengths. Here we report on TDLAS- and OES-based measurements of modal temperatures and species number density time-histories of NO, O, and N in the post-incident-shock region of NASA EAST experiments. Thus, this section describes the EAST facility setup, provides background for the absorption and emission processes, and specifies the design and operation of the respective TDLAS and OES diagnostics.

\subsection{Shock Tube Setup}

\begin{figure}[htb!]
\centering
\includegraphics[width=.98\textwidth, trim = {0cm, 0cm, 0cm, 0cm}, clip]{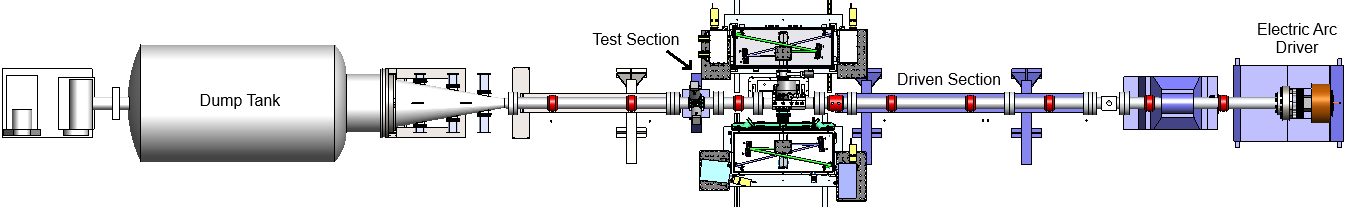}
\caption{\label{fig:EAST} Top view of the EAST HVST, showing the electric arc driver, 10.16 cm tube diameter and OES setup coupled to the test section. TDLAS integration is shown in Fig. \ref{fig:setup}. Adapted from Refs. \cite{Fagnani2025, Fagnani2026}, with permission.}
\end{figure}

All measurements presented here were collected at the EAST High Velocity Shock Tube (HVST). The HVST is a 10.16 cm diameter, arc-driven aluminum shock tube; Fig.~\ref{fig:EAST} presents a top-view schematic of the facility. The driven section is filled with the test gas -- in this case hydrocarbon-free air, treated as 21\% O$_2$ with balance N$_2$ for modeling \cite{cruden2024b}. An electric current is discharged through the driver gas (helium/argon) to generate a large increase in the driver chamber pressure and temperature, breaking a diaphragm and leading to formation of a propagating shock wave in the driven tube. The use of the electric arc driver can produce a wide range of shock velocities, including the very high velocities relevant to atmospheric entry. The test section begins 7.5 m downstream of the diaphragm and is instrumented with ports and windows for optical access (TDLAS and emission sensors) and time-of-arrival sensors. TDLAS lines-of-sight are spaced over the same region as the 13.4 cm slot window used for spatially-resolved emission measurements. TDLAS lines-of-sight could only co-locate two lasers at each location, so the three NO laser measurements and four O/N measurements were located between 7.82 m and 7.95 m downstream of the diaphragm. Time-of-arrival sensors are placed 3.175 cm apart for high-resolution velocity measurements, as the local shock velocity is required for defining the initial post-shock conditions. Additional facility details may be found in \cite{east}, and the integration of the laser diagnostics with EAST is further described in Section \ref{diagnostics}. Table \ref{Conditions} provides reference numbers, fill pressures, and shock velocities for the experimental conditions investigated.

\setcounter{table}{0}
\begin{table}[htb!]
\centering
\caption{Summary of the shock conditions analyzed in this work.}\label{Conditions}\vspace{-3mm}
\begin{tabular}{ c c c }
  \hline \hline
    \textbf{Shot ID} & \textbf{Fill pressure} & \textbf{Shock velocity} \\ 
      & Torr & km/s \\  
      \hline
    66/21 & 9.96 & 3.98 \\  
    66/26 & 2.25 & 3.81 \\
    66/41 & 1.40 & 7.03 \\
    \hline \hline
\end{tabular}
\end{table}

%=======================================================================================================
\subsection{Tunable Diode Laser Absorption Spectroscopy}
\subsubsection{Scanned Direct Absorption} 
Spectroscopic techniques including TDLAS have been applied as a non-intrusive flow diagnostic in various hypersonic ground test facilities, providing species and state-specific information from which additional flow properties are inferred \cite{las}. Furthermore, TDLAS has been applied on EAST before, with past applications probing CO and CO$_2$ \cite{Jelloian_EAST_2022,MacDonald_EAST_2016} as opposed to the current application to NO, O, and N. As with past TDLAS systems, current work implements scanned direct absorption (SDA), an application of TDLAS in which a laser is current-injection-tuned across the absorbing bandwidth of one or more quantum transitions.
For any wavenumber output of the tunable diode laser, the spectral absorbance $\alpha_\nu$ at wavenumber $\nu$ can be calculated from the fractional transmission of light $\frac{I_t}{I_0}$ (Eqn. \ref{eq:beer}). Absorbance may then be related to the pressure $P$, optical path length $l_{abs}$, mole fraction of the absorbing species $X_{\text{abs}}$, temperature-dependent linestrength of the quantum transition $S_k(T)$, and the lineshape function of the quantum transition $\phi_k(\nu)$, with total absorbance arising from the summed contribution from all overlapping absorption features, k.
As the EAST flows are expected to be uniform in the plane of the laser diagnostics, and boundary layers remain small for the current conditions, $l_{abs}$ corresponds to the EAST internal diameter (D = 10.16 cm).
For SDA, frequency integration across the scanned absorbance feature eliminates the dependence on the lineshape $\phi$ in Eqn. \ref{eq:beer}, as the integration of $\phi$ by definition is equal to one.

\begingroup
\small
\begin{equation}\label{eq:beer}
\begin{aligned}
    \alpha_\nu=-\ln\left(\frac{I_t}{I_0}\right) = \sum_kPX_{\text{abs}}S_k(T)\phi_{k}(\nu)l_{abs}
\end{aligned}
\end{equation}
\endgroup  

\begingroup
\small
\begin{equation}\label{eq:intA}
    A_k = PX_{\mathrm{abs}}S_k(T)l_{abs}
\end{equation}
\endgroup

The integration in frequency space then yields Eqn. \ref{eq:intA}, which enables inferences of species number density after applying a Voigt fit to absorbance data. The Voigt profile is defined by a convolution of a Gaussian and Lorentzian distribution to capture both Doppler ($\Delta \nu_D$) and collisional ($\Delta \nu_C$) broadening, respectively.
For a single, well-isolated transition, a Doppler-temperature may also be extracted from Gaussian component of the Voigt lineshape data (i.e. $\Delta \nu_D$), according to Eqn. (\ref{eq:doppler}). Pressure may also be extracted from the Lorentzian component of the Voigt lineshape data (i.e. $\Delta \nu_C$), although this was not pursued for the current measurements.

\begingroup
\small
\begin{equation}\label{eq:doppler}
    \Delta \nu_{\textrm{D}} = 7.1623\times 10^{-7}\nu_{0,k}\sqrt{\frac{T_{trans}}{M}} 
\end{equation}
\endgroup

Inferences of Doppler-temperature were pursued for the low-pressure, high-temperature experiments probing atomic species, while modal temperatures  of NO were inferred from integrated area values.
When multiple transitions of differing lower vibrational (v") and rotational (J") states are captured for a single species, the modal temperatures may be separately resolved by expressing Eqn. \ref{eq:intA} in terms of spectroscopic constants, as in Eqn. \ref{eq:4}. This formulation assumes a Boltzmann distribution between the upper (j) and lower (i) states at temperature $T$, as made explicit in Eqn.  \ref{eq:4}.
\begingroup
\small
\begin{equation}\label{eq:4}
    A_k = l_{abs}\frac{n_i\mathcal{A}_{ji}g_{j}}{8\uppi \speedoflight \nu_{0,k}^2g_{i}}\left(1-\left(\frac{n_j/g_j}{n_i/g_i}\right)\right) = l_{abs}\frac{n_i\mathcal{A}_{ji}g_{j}}{8\uppi \speedoflight \nu_{0,k}^2g_{i}}\left(1-\mathrm{exp}\left(-\frac{h\speedoflight\nu_{0,k}}{\boltzmannconstant T}\right)\right)
\end{equation}
\endgroup

Then, expanding for distinct rotational and vibrational distributions in the case of NO, 
\begingroup
\small
\begin{equation}\label{eq:6}
    A_k = n_{total}^sl_{abs}\frac{\mathcal{A}_{ji}g_{j}}{8\uppi \speedoflight \nu_{0,k}^2 Q(\Trot,\Tvib)} \mathrm{exp}\left(-\left(\frac{E_{\textrm{rot},k}}{\boltzmannconstant \Trot }+\frac{E_{\textrm{vib},k}}{\boltzmannconstant \Tvib}\right)\right)\left(1-\mathrm{exp}\left(-\left(\frac{\Delta E_{\textrm{rot},k}}{\boltzmannconstant \Trot}+\frac{\Delta E_{\textrm{vib},k}}{\boltzmannconstant \Tvib}\right)\right)\right).
\end{equation}
\endgroup

Note that the Boltzmann distribution assumption in Eqn. \ref{eq:4} only applies to the upper and lower states, while the expansion to Eqn. \ref{eq:6} adds an additional assumption that the rotational and vibrational states are Boltzmann distributed across all states -- even those not directly involved in the absorption transition. This secondary Boltzmann assumption arises from the Boltzmann population fraction assumption inherent in expressing the absorbing state number density ($n_i$) in terms of overall number density ($n_{total}^s$), modal temperatures ($\Trot$ and $\Tvib$), and partition function ($Q$). This second Boltzmann assumption could be relaxed via a non-Boltzmann partition function, although such a correction would not substantially alter the inferred n$_{total}^{NO}$, $T_{\mathrm{rot}}$, or $T_{\mathrm{vib}}$ values for the current NO measurements.

The spectroscopic constants $\mathcal{A}_{ji}$, $g_{j}$, $g_{i}$, $\nu_{0,k}$, and $E_{rot/vib,k}$ are obtained from the HITEMP database for NO \cite{hitemp}, and overlapping states due to spin splitting and lambda doubling are included in the modeling of A$_k$. The total partition function $Q$ is calculated by summing contributions from all rotational and vibrational levels, calculated with rovibrational energy level parameters from \cite{AMIOT1982150}.
For the selected NO absorption features, these spectroscopic and energy parameters are summarized in Table \ref{tab:NOlines}.
For NO, the measured absorbance data are then fit to Eqn. (\ref{eq:6}) with a weighted least-squares fitting scheme according to Eqn. (\ref{eq:weightscheme}), where $A_{\mathrm{model},k}$ refers to the simulated integrated absorbance for transition k, $A_{\mathrm{meas},k}$ to the corresponding TDLAS measurement, and $w_k$ an uncertainty weight so that measurements with higher uncertainty are weighted less heavily.

\begingroup
\small
\begin{equation}\label{eq:weightscheme}
   \arg\min\left(\sum_m\left(\begin{bmatrix}
    A_{\mathrm{model,1}}(n_{total}^s,T_{\mathrm{rot}},T_{\mathrm{vib}})\\
    \vdots\\
    A_{\mathrm{model,m}}(n_{total}^s,T_{\mathrm{rot}},T_{\mathrm{vib}})\\
\end{bmatrix} - 
\begin{bmatrix}
    A_{\mathrm{meas,1}}\\
    \vdots\\
    A_{\mathrm{meas,m}}\\
\end{bmatrix}\right)^2 \cdot
\begin{bmatrix}
    w_1\\
    \vdots\\
    w_m\\
\end{bmatrix}\right)
\end{equation}
\endgroup

In contrast to NO where all states were fit subject to Eqns. \ref{eq:4}-\ref{eq:weightscheme}, the O and N states were only analyzed subject to Eqns. \ref{eq:doppler} and \ref{eq:4}. Eqn. \ref{eq:doppler} was used to infer a Doppler-temperature from the Gaussian portion of the lineshape profile, while Eqn. \ref{eq:4} was sufficient for inferring the absorbing state number density, $n_i$ for each probed state. For these transitions, spectroscopic constants are taken from the NIST Atomic Spectra Database (ASD) for atomic species, and reported in Table \ref{tab:lines}. Where muliplet transitions overlap, multiple overlapping Voigt profiles were used to improve the fits, with the relative linecenter positions of the overlapping transitions constrained by the NIST ASD values.

Overall, the SDA measurements thus enabled state-specific number density and modal temperature time-histories to be inferred from the measured absorbance values.
In each case, the specific absorbing states were determined through the wavelength selection (Section \ref{selection}), while the modal temperatures differed between the NO measurements and the atomic state measurements.
For NO, the two inferred modal temperatures were the rotational and vibrational temperatures ($T_{\mathrm{rot}}$ and $T_{\mathrm{vib}}$, respectively), which were determined through the Boltzmann population distributions of the absorbing states.
In contrast, the O- and N-atom measurements inferred two other modal temperatures -- namely translational temperature from the Doppler widths ($T_{\mathrm{trans}}$) and the electronic temperature through an assumed Boltzmann population distribution of the electronic states ($T_{\mathrm{elec}}$).
Combined, these varied measurements demonstrate the value of SDA for rapid, sensitive inferences of state-specific data in high-temperature, nonequilibrium air.

\subsubsection{TDLAS Wavelength Selection}\label{selection}
For the slower incident shock velocities (<5 km/s), NO was selected as the target species due to demonstrated success in similar hypersonic applications, as well as a  well-understood, easily probed IR absorbance spectrum \cite{caltech,almodovar,spearrin,varghese}. The current work advances upon prior studies by demonstrating multi-temperature sensing capabilities at even higher temperatures via SDA, spectrally resolving several rovibrational lines in the NO fundamental vibrational bands. Specific transitions were selected based on the following criteria:
\begin{enumerate}
    \item Spectral isolation from interfering species, including ambient water and carbon dioxide, for interference-free detection,
    \item Strong absorbance (>0.02) to maximize SNR, 
    \item A maximum number and range of lower-state quanta (v", J") for enhanced temperature sensitivity, and
    \item Groups of quantum transitions that can be captured in a single scan at >100 kHz scan rates, depending on the scan rates and depths available for typical TDLAS systems, for sensor compactness and short test time considerations.
\end{enumerate}

\begin{figure}[htb!]
\centering
\includegraphics[width=.8\textwidth]{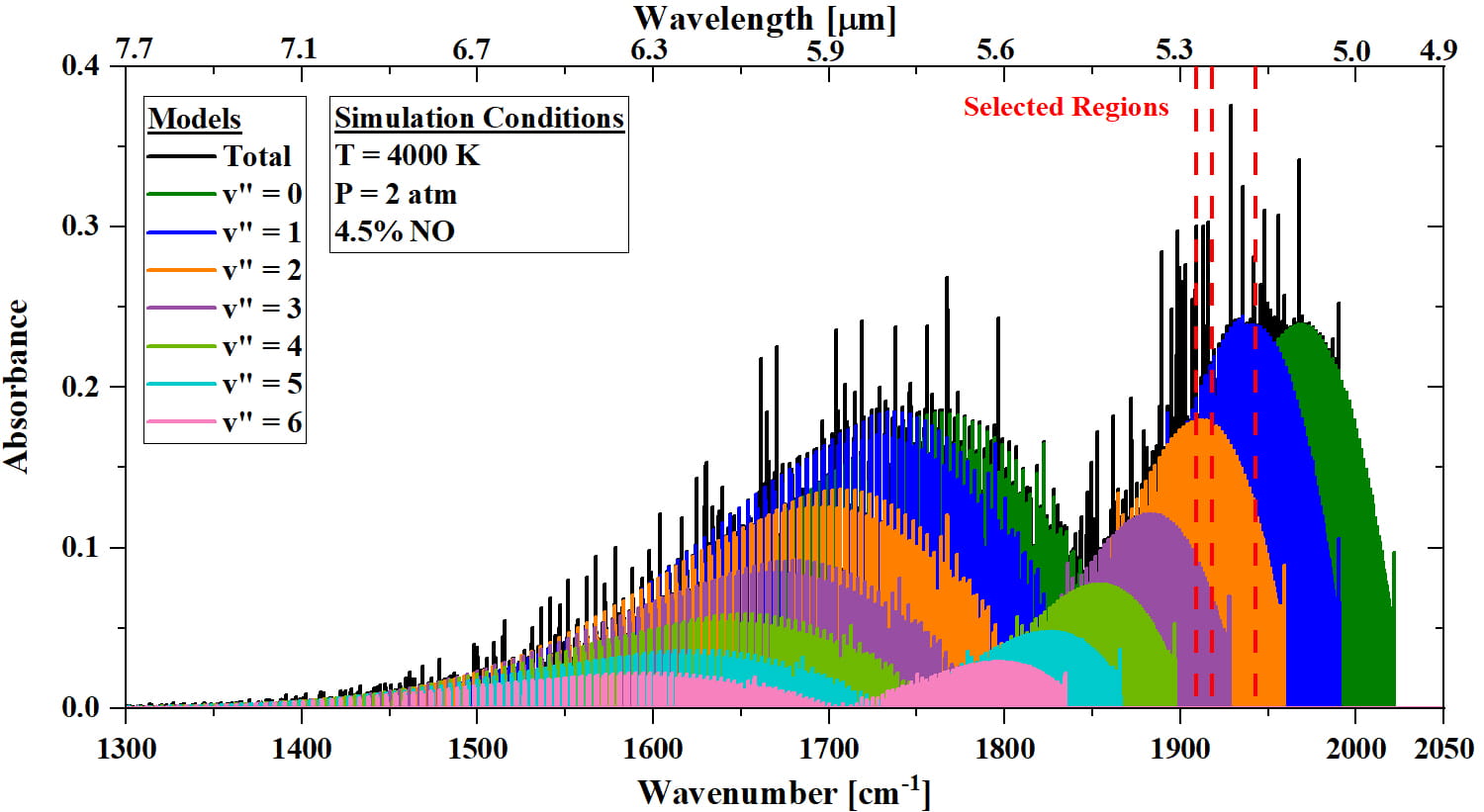}
\caption{\label{fig:abs} HITEMP \cite{hitemp} simulated absorbance of NO fundamental bands, with selected wavelengths shown via red dashed lines. Colors show vibrational state content, with total absorbance in black.}
\end{figure} 

\begin{figure}[htb!]
\centering
\subfigure[]
{{\includegraphics[width=.49\textwidth]{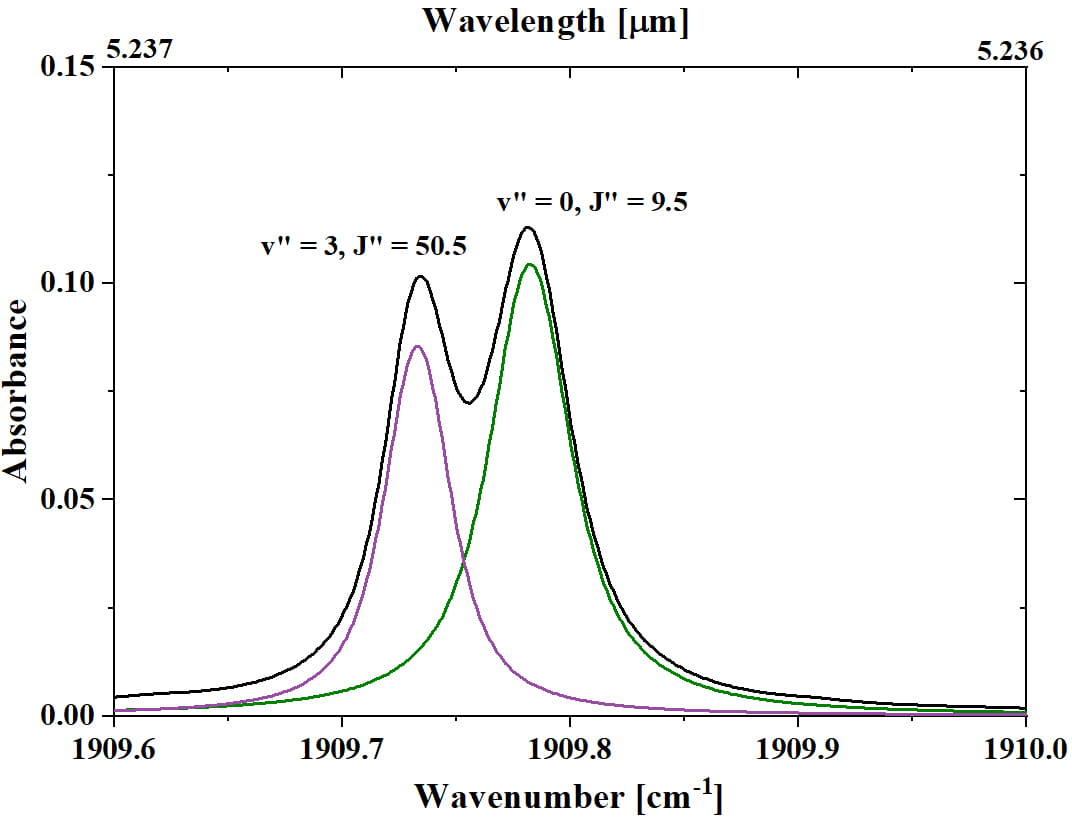}}}%
\subfigure[]
{{\includegraphics[width=.49\textwidth]{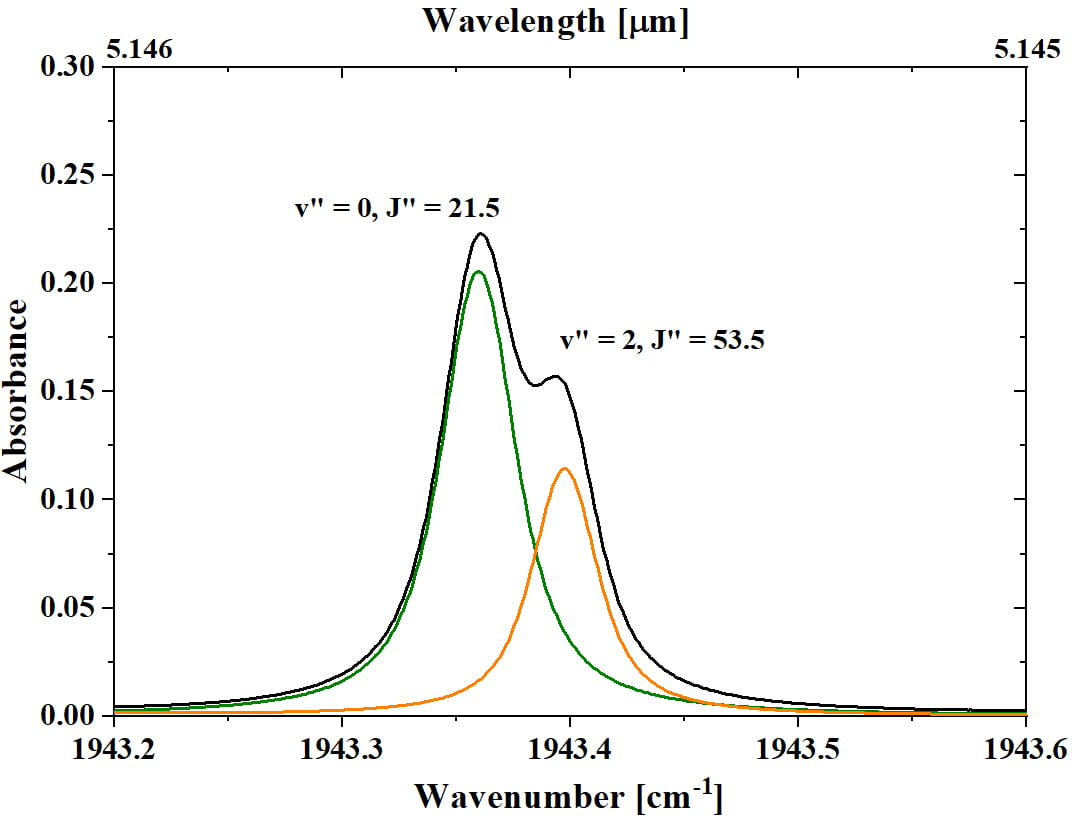}}}\\
\subfigure[]
{{\includegraphics[width=.49\textwidth]{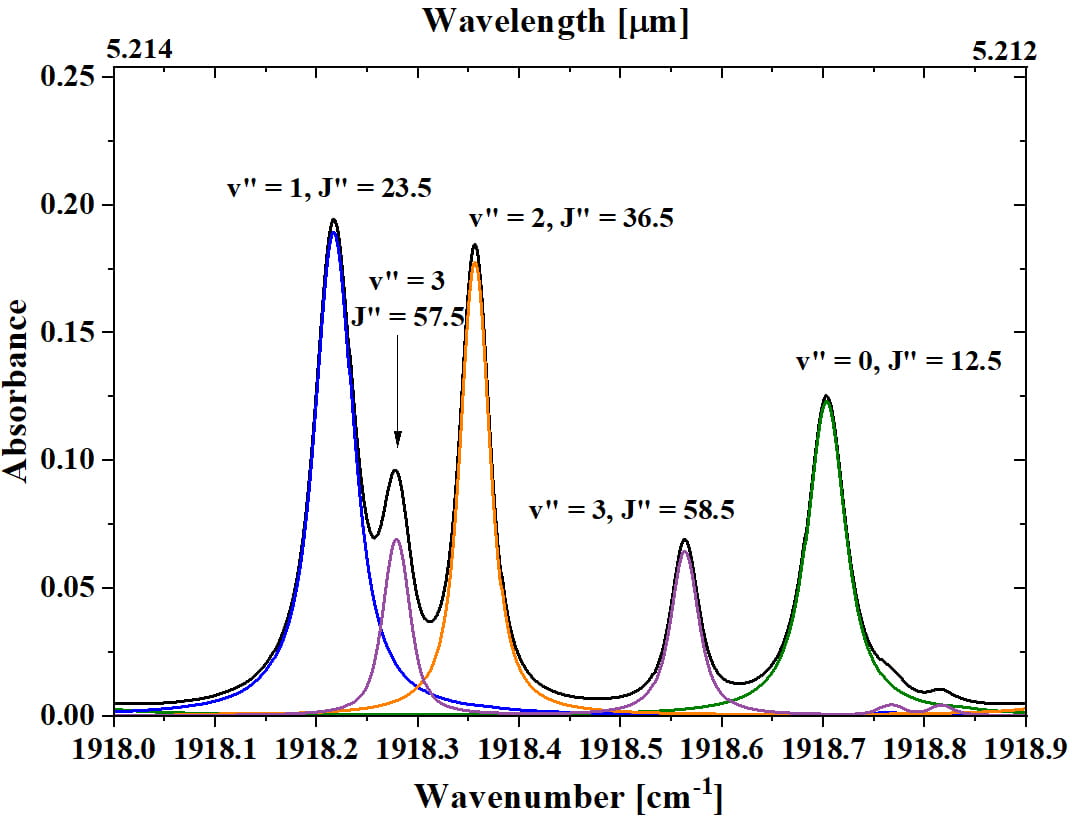}}}%
\caption{\label{fig:abs_individual}Simulated absorbance for selected wavelength groups at the same $ T = \SI{4000}{\kelvin}$, $P = \SI{2}{atm}$, 4.5\% NO condition as Fig. \ref{fig:abs}; (a) 1909 cm$^{-1}$ region; (b) 1943 cm$^{-1}$ region; (c) 1918 cm$^{-1}$ region.}
\end{figure}

The resulting NO wavelength selection utilized three lasers, covering nine features around $\SI{5.2}{\micro \meter}$ with v" = 0-3 and J" = 9.5-58.5. 
Figure \ref{fig:abs} presents the NO absorbance spectrum at a representative test condition, with path length set to the D = 10.16 cm inner diameter of EAST. Figure \ref{fig:abs_individual} presents the selected wavelength groups at the same condition. Finally, Table \ref{tab:NOlines} provides the data from the HITEMP database that were used for inferences of modal temperature and NO number densities \cite{hitemp}. As mentioned, overlapped states of NO arise from lambda doubling - denoted with the e/f notation. However, HITEMP further separates hyperfine splitting of lines in v" = 0 and 1 states due to their correspondingly lower J" values, resulting in six entries for those states. For states with v" > 1 and relatively high J" values, the hyperfine splitting is grouped together in the HITEMP database, resulting in single entries with correspondingly larger degeneracies. Regardless, the overlapped states were considered when modeling A$_k$ by summing either the two or six entries from HITEMP (Table \ref{tab:NOlines}). Note that HITEMP does not separate $E_{\textrm{rot}}$ and $E_{\textrm{vib}}$, so these values (along with $\Delta E_{\textrm{rot}}$ and $\Delta E_{\textrm{vib}}$) were calculated using Dunham coefficients from Amiot \cite{AMIOT1982150}. However, the sum of $E_{\textrm{rot}}$ and $E_{\textrm{vib}}$ was confirmed to be equal to the total lower-state energy, as provided in the HITEMP database. The upper-state quanta (v', J'), used to calculate  $\Delta E_{\textrm{rot}}$ and $\Delta E_{\textrm{vib}}$, were easily determined as $\Delta$v = v'-v" = + 1 for all absorption transitions in the fundamental band, while the chosen R-branch features correspond to $\Delta$J = J'-J" = + 1. Finally, the partition function used was adapted from one generated for a recent spectroscopic model of UV NO absorbance \cite{Ajay3}.

\setlength{\tabcolsep}{3pt}
\begin{table}
\centering
\renewcommand{\arraystretch}{1.5}
\caption{NO molecular transitions considered in this work, enabling measurements of NO modal temperatures and number density.\cite{hitemp,AMIOT1982150,Ajay3} All transitions are R-branch, meaning $\Delta$J = J'-J" = + 1.}\label{tab:NOlines}\vspace{-3mm}
\small
%\resizebox{\textwidth}{!}{
\begin{tabular}{cccccccccccc}
\hline \hline
 \textbf{Vib. State} & \textbf{Rot. State} &\textbf{ Parity} &\textbf{ Hyperfine} & \textbf{Wavenumber} & $\mathbf{g_i}$ & $\mathbf{g_j}$ & $\mathbf{\mathcal{A}_{ji}}$ & $\mathbf{E_{rot}}$ & $\mathbf{E_{vib}}$ & $\mathbf{\Delta E_{rot}}$ & $\mathbf{\Delta E_{vib}}$ \\
 v" & J" &  & F" & $\nu_{0,k}$, cm$^{-1}$ &  &  & $\mathrm{s^{-1}}$ & $\mathrm{cm^{-1}}$ & $\mathrm{cm^{-1}}$ & $\mathrm{cm^{-1}}$ & $\mathrm{cm^{-1}}$ \\
 \hline
\multirow{2}{*}{3} & \multirow{2}{*}{50.5} &  e &      & 1909.73248 & 306 & 312 & 24.82 &  4193.17 & \multirow{2}{*}{5543.73} & \multirow{2}{*}{118.44} & \multirow{2}{*}{1791.85} \\ \cline{3-9}
                   &                       &  f &      & 1909.73454 & 306 & 312 & 24.82 &  4193.45 &                          &                         &                          \\ \hline
\multirow{6}{*}{0} & \multirow{6}{*}{9.5}  & \multirow{3}{*}{e}    & 10.5 & 1909.78196 & 22  & 24  & 6.06  & \multirow{3}{*}{289.95}     & \multirow{6}{*}{0.00}    & \multirow{6}{*}{33.36}  & \multirow{6}{*}{1875.99} \\
                   &                       &                       & 9.5  & 1909.78202 & 20  & 22  & 6.01  &                             &                          &                         &                          \\
                   &                       &                       & 8.5  & 1909.78207 & 18  & 20  & 6.00  &                             &                          &                         &                          \\ \cline{3-9}
                   &                       & \multirow{3}{*}{f}    & 10.5 & 1909.78338 & 22  & 24  & 6.06  & \multirow{3}{*}{289.96}     &                          &                         &                          \\
                   &                       &                       & 9.5  & 1909.78343 & 20  & 22  & 6.01  &                             &                          &                         &                          \\
                   &                       &                       & 8.5  & 1909.78347 & 18  & 20  & 6.00  &                             &                          &                         &                          \\ \hline \hline
\multirow{6}{*}{0} & \multirow{6}{*}{21.5} & \multirow{3}{*}{e}    & 22.5 & 1943.35775 & 46  & 48  & 6.65  & \multirow{3}{*}{948.31}     & \multirow{6}{*}{0.00}    & \multirow{6}{*}{66.67}  & \multirow{6}{*}{1875.99} \\
                   &                       &                       & 21.5 & 1943.35778 & 44  & 46  & 6.63  &                             &                          &                         &                          \\
                   &                       &                       & 20.5 & 1943.35781 & 42  & 44  & 6.63  &                             &                          &                         &                          \\ \cline{3-9}
                   &                       & \multirow{3}{*}{f}    & 22.5 & 1943.36241 & 46  & 48  & 6.65  & \multirow{3}{*}{948.35}     &                          &                         &                          \\
                   &                       &                       & 21.5 & 1943.36241 & 44  & 46  & 6.63  &                             &                          &                         &                          \\
                   &                       &                       & 20.5 & 1943.36242 & 42  & 44  & 6.63  &                             &                          &                         &                          \\ \hline
\multirow{2}{*}{2} & \multirow{2}{*}{53.5} &  e &      & 1943.39719 & 324 & 330 & 19.89 &  4748.11 & \multirow{2}{*}{3723.88} & \multirow{2}{*}{124.08} & \multirow{2}{*}{1819.85} \\ \cline{3-9}
                   &                       &  f &      & 1943.39880 & 324 & 330 & 19.88 &  4748.38 &                          &                         &                          \\ \hline \hline
\multirow{6}{*}{1} & \multirow{6}{*}{23.5} & \multirow{3}{*}{e}    & 22.5 & 1918.21421 & 46  & 48  & 12.94 & \multirow{3}{*}{951.07}     & \multirow{6}{*}{1875.98} & \multirow{6}{*}{71.02}  & \multirow{6}{*}{1847.89} \\
                   &                       &                       & 23.5 & 1918.21422 & 48  & 50  & 12.95 &                             &                          &                         &                          \\
                   &                       &                       & 24.5 & 1918.21424 & 50  & 52  & 12.97 &                             &                          &                         &                          \\ \cline{3-9}
                   &                       & \multirow{3}{*}{f}    & 24.5 & 1918.22071 & 50  & 52  & 12.97 & \multirow{3}{*}{951.30}     &                          &                         &                          \\
                   &                       &                       & 22.5 & 1918.22071 & 46  & 48  & 12.95 &                             &                          &                         &                          \\
                   &                       &                       & 23.5 & 1918.22072 & 48  & 50  & 12.95 &                             &                          &                         &                          \\ \hline
\multirow{2}{*}{3} & \multirow{2}{*}{57.5} &  e &      & 1918.27432 & 348 & 354 & 25.19 &  5634.23 & \multirow{2}{*}{5543.73} & \multirow{2}{*}{125.96} & \multirow{2}{*}{1791.85} \\ \cline{3-9}
                   &                       &  f &      & 1918.28382 & 348 & 354 & 25.19 &  5634.65 &                          &                         &                          \\ \hline
\multirow{2}{*}{2} & \multirow{2}{*}{36.5} &  e &      & 1918.35280 & 222 & 228 & 18.97 &  2406.93 & \multirow{2}{*}{3723.88} & \multirow{2}{*}{97.79}  & \multirow{2}{*}{1819.85} \\ \cline{3-9}
                   &                       &  f &      & 1918.36025 & 222 & 228 & 18.97 &  2407.09 &                          &                         &                          \\ \hline
\multirow{2}{*}{3} & \multirow{2}{*}{58.5} &  e &      & 1918.56290 & 354 & 360 & 25.21 &  5599.45 & \multirow{2}{*}{5543.73} & \multirow{2}{*}{127.16} & \multirow{2}{*}{1791.85} \\ \cline{3-9}
                   &                       &  f &      & 1918.56425 & 354 & 360 & 25.21 &  5599.71 &                          &                         &                          \\ \hline
\multirow{6}{*}{0} & \multirow{6}{*}{12.5} & \multirow{3}{*}{e}    & 13.5 & 1918.70212 & 28  & 30  & 6.26  & \multirow{3}{*}{408.46}     & \multirow{6}{*}{0.00}    & \multirow{6}{*}{42.19}  & \multirow{6}{*}{1875.99} \\
                   &                       &                       & 12.5 & 1918.70217 & 26  & 28  & 6.23  &                             &                          &                         &                          \\
                   &                       &                       & 11.5 & 1918.70220 & 24  & 26  & 6.22  &                             &                          &                         &                          \\ \cline{3-9}
                   &                       & \multirow{3}{*}{f}    & 13.5 & 1918.70430 & 28  & 30  & 6.26  & \multirow{3}{*}{408.47}     &                          &                         &                          \\
                   &                       &                       & 12.5 & 1918.70433 & 26  & 28  & 6.23  &                             &                          &                         &                          \\
                   &                       &                       & 11.5 & 1918.70436 & 24  & 26  & 6.22  &                             &                          &                         &                           \\
 \hline \hline
\end{tabular}
%}
\end{table}

\begin{figure}[htb!]
\centering
\subfigure[]
{{\includegraphics[width=.49\textwidth]{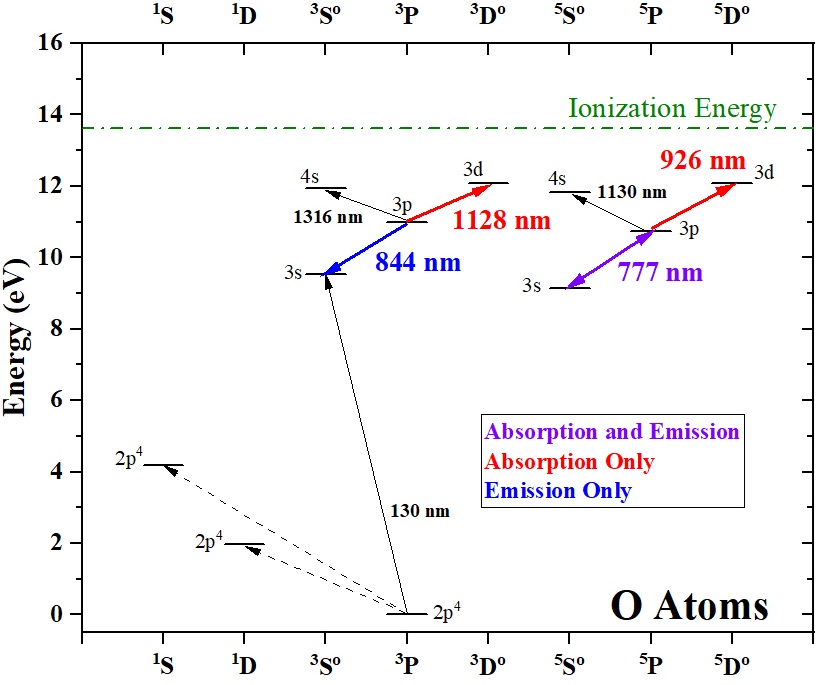}}}%
\subfigure[]
{{\includegraphics[width=0.49\textwidth]{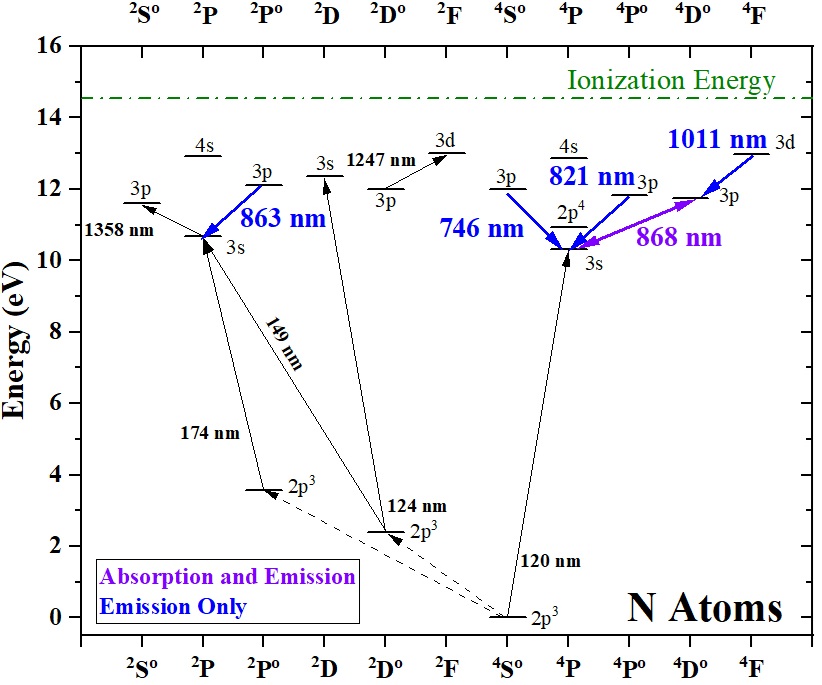}}}%
\caption{\label{fig:energylvls}Simplified energy level diagrams for (a) atomic oxygen, and (b) atomic nitrogen. Term symbols are denoted along the x-axis and electronic configuration next to corresponding levels.}
\end{figure}

\newpage
At the higher incident shock velocities (>5 km/s) where the molecular nitrogen and oxygen dissociate into atomic N and O, radiation can be traced to ionized species or excited quantum states of neutral species. At these conditions, O and N were selected as the target species. The wavelength selection for these is straightforward due to the limited set of electronic transitions for atomic species. Energy level diagrams for both species are provided in Fig \ref{fig:energylvls} with corresponding wavelength options. The selected states are fairly high in energy, but are easily accessible for probing in the NIR. By contrast, the ground and metastable states are only accessible through vacuum ultraviolet (VUV) absorption \cite{Granowitz2025}. The resulting excited species wavelength selection covers the single O transition near 777.2 nm ($\mathrm{3s^5S^{\circ}_2\rightarrow3p^5P_3}$), three transitions near 926.6 nm ($\mathrm{3p^5P_3\rightarrow3d^5D^{\circ}_{2,3,4}}$), six transitions near 1128.7 nm ($\mathrm{3p^3P_{0,1,2}\rightarrow3d^3D^{\circ}_{1,2,3}}$), and the single N transition near 868.0 nm ($\mathrm{3s^4P_{5/2}\rightarrow3d^4D^{\circ}_{7/2}}$) \cite{yang,peter}. See Table \ref{tab:lines} for additional details of the atomic state transitions.

\subsubsection{Laser Diagnostics}\label{diagnostics}
\begin{figure}[hbt!]
\centering
\includegraphics[width=.95\textwidth]{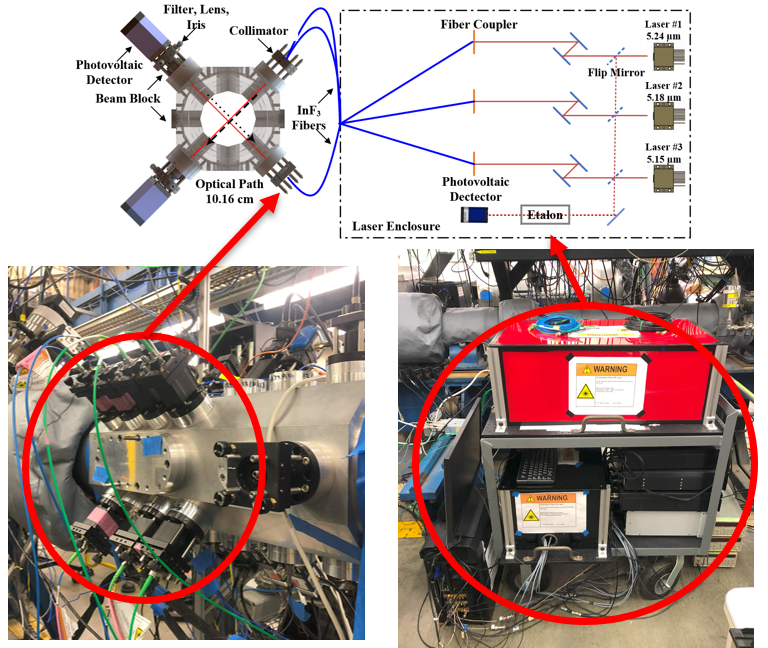}
\caption{\label{fig:setup}The alignment (Top), integration with EAST (Bottom Left), and laser enclosure (Bottom Right) for the TDLAS sensor diagnostics. Reproduced from Ref. \cite{Changthesis}, with permission.}
\end{figure}

Sensor architecture was optimized for portability and integration. Both the MIR (NO) and NIR (O, N) sensor systems were assembled to fit on individual 60x90 cm breadboards and were fully cage-mounted, light-tight, and fiber-coupled for ease of use and safety. Figure \ref{fig:setup} is a schematic of the NO sensor architecture and integration with the EAST facility. Corresponding photos are provided directly below, with MIR NO laser components housed in the red enclosure and NIR O and N laser components in the black enclosure. The internal alignment (Fig. \ref{fig:setup} Top) and integration with EAST is identical for the NIR sensor (O, N) with optics and equipment swapped appropriately for wavelength. In this work, all lasers were distributed feedback (DFB) lasers, covering wavelengths around $\SI{5.2}{\micro \meter}$ for NO, and 777.2 nm, 926.6 nm, and 1128.7 nm for O, and 868.0 nm for N. Each laser was current-injection-tuned with a sinusoidal waveform over a narrow range of wavelengths for the SDA probing of lineshapes and linestrengths of specified quantum transitions. The optical alignment was optimized to mitigate non-ideal effects including density-gradient beam steering and emission. 

For a given laser, the laser scan rate is a balance between temporal resolution and spectral scan range, directly dependent on and inversely dependent on scan rate, respectively. Hence, all lasers were modulated at 500 kHz, to maintain relatively high temporal resolution. However, each laser system also exhibits a phase shift between the intensity modulation and the wavelength output, which increases for fast scan rates. 
Thus, absorption features tend to fall on portions of the scan with higher intensity on the upscan (increasing output intensity) than their corresponding location on the downscan (decreasing output intensity).
When both upscan and downscan are well-resolved, the measurement rate is double that of the scan frequency, allowing measurements at 1 MHz.
However, some lasers (e.g. the 1918 cm$^{-1}$ and 1128.7 nm laser systems) exhibit so substantial of a phase shift that the downscan data is either poorly resolved or shifted to regions of no output intensity.
Thus, select measurements were only obtained for upscans, reducing their temporal resolution to 500 kHz.
To clarify these considerations, the TDLAS results will use solid symbols for upscan and open symbols for downscan data throughout this work.
These phase-shift and scan depth considerations highlight the laser architecture considerations from criteria 4) in Section \ref{selection}.

\par The laser beams were passed through 2$^{\circ}$ wedged CaF$_2$ or Al$_2$O$_3$ (sapphire) windows, and collected by 220 MHz high-bandwidth Vigo Systems PV detectors with 1 mm$^2$ active area (for NO) or by 150 MHz high-bandwidth Thorlabs PDA10A2 detectors with 0.8 mm$^2$ active area (for O at 777 nm and 926 nm and for N at 868 nm). Additionally, a 150 MHz high-bandwidth Thorlabs PDA10CF detector with 0.2 mm$^2$ active area was used for O at 1128 nm. Focusing lenses, 10 nm half bandwidth spectral filters, and neutral density (ND) filters were inserted into the detector cage system to further optimize signal and minimize emission interference. Detector signals were recorded on a Gagescope system with a 12-bit depth at a rate of 3 GS/s. A custom-coated YAG-etalon with a free spectral range (FSR) of 0.0272 cm$^{-1}$ was used in the NIR and a Ge-etalon with FSR = 0.0162 cm$^{-1}$ used in the MIR to obtain a time-to-frequency conversion of the laser light. For the NO measurements (shots 66/21 and 66/26), the 1909 cm$^{-1}$ and 1919 cm$^{-1}$ lasers were co-located 7.89 m downstream of the diaphragm, while 1943 cm$^{-1}$ was located 7.858 m downstream. For the O and N measurements (shot 66/41), 777 nm and 868 nm were located 7.953 m downstream, 926 nm was located 7.89 m downstream, and 1128 nm was located 7.826 m downstream of the diaphragm. Together, these sensitive diagnostic equipment, alignment strategies, and integration with the EAST facilities enabled the high-speed measurements of NO, O, and N absorption transitions.

%=======================================================================================================
\subsection{Optical Emission Spectroscopy}
\subsubsection{Spatially and Spectrally Resolved Emission Measurements}

The radiative emission from excited states was probed using four spectrometers spanning the UV to the NIR or MIR wavelength range, and previously analyzed for different conditions achieved in Test Series 66 by \citet{cruden2024b}. A subset of the collected data are analyzed in this work, with focus on the atomic line emission intensities for comparison with the TDLAS measurements.

Calibrated OES frames map the radiance emitted across the tube diameter, resolved in wavelength, and spatial position along the view window. Table~\ref{tab:spectrometers} provides details on the spectrometers used for the conditions considered in this paper. 
The intensity calibration of the OES measurements follows the methodology discussed by~\citet{cruden2014a}, and further detailed here. Assuming linearity between the dark-subtracted digitalized intensity counts, $\hat{U}_p = U_p - U_{p, \textrm{bg}} $, and the incident radiance $L_{\lambda_p}$ onto the detector pixel $p$, the radiometric measurement equation gives
\begin{equation}
	\hat{U}_p  = \alpha\subrm{det} \times \Theta \times \mathcal{R}_{\lambda_p} \times  \int_{\Delta x_p} \int_{\Delta \lambda_p} \int_{\Delta t} f\subrm{gate}(t) L_{\lambda_p}(\lambda, x, t)  dt d\lambda dx,
	\label{eqn:OES_meas}
\end{equation}
where $ \alpha\subrm{det}$ is the detector constant, $\Theta$ is the instrument throughput, and $\mathcal{R}_{\lambda_p}$ is the absolute spectral radiance responsivity.  $ \Delta\lambda_p $ and $ \Delta x_p$ are the spectral and spatial dimensions covered by the detector pixel, respectively, while $\Delta t$ is the detector gate time. $f\subrm{gate}$ represents the detector gate function, which results in a modulation of the signal, particularly at short gate times.
Considering the integral average
\begin{equation}
	\bar{L}_{\lambda_p} = \frac{1}{\Delta x_p} \frac{1}{\Delta \lambda_p} \frac{r\subrm{gate}(\Delta t)}{\Delta t} \int_{\Delta x_p} \int_{\Delta \lambda_p} \int_{\Delta t} f\subrm{gate}(t) L(\lambda, x, t) d\lambda dx dt,
	\label{eqn:OES_Lavg}
\end{equation}
where $r\subrm{gate}$ is the gating reciprocity factor, and applying Eqns.~\ref{eqn:OES_meas} and~\ref{eqn:OES_Lavg} to both measurement and calibration frames, indicated with the superscripts ``m'' and ``c'', respectively, yields
\begin{equation}
	\begin{cases}
		\bar{L}\subsuprm{\lambda_p}{m} = \frac{\hat{U}_{p, m}}{F_{\lambda_p}/R} \\
		F_{\lambda_p}/R = \frac{\hat{U}_{p, c}}{\bar{L}\subsuprm{\lambda_p}{c}} \times \frac{r\subrm{c}/\Delta t\subrm{c}}{r\subrm{m}/\Delta t\subrm{m}},
	\end{cases}
    \label{eqn:OES_calib}
\end{equation}
where $F_{\lambda_p}$ indicates the sensor response, and $R$ is the reciprocity factor ratio.

\begin{table}
\caption{Details of the emission spectrometer parameters employed for this study.}\vspace{-6mm}
\label{tab:spectrometers}
\begin{center}
\begin{tabular}{ccccc}
\hline
\hline
\textbf{Spectral Region}&\textbf{Spectrometer} &\textbf{Camera} &\textbf{Grating [grooves/mm]} &\textbf{Range [nm]} \\
\hline
Red & \multirow{3}{*}{SP2500}   & PIMAX2 & 50 & 500-880\\
NIR &                           & NIT (WiDy SenS 640) & 30 & 970-1550\\
MIR &                            & IRC800 & 17 & 5000-5700\\
\hline
\hline
\end{tabular}
\end{center}
\end{table}

The calibration sources employed at EAST include a 4" aperture integrating sphere (Labshere Inc, NH, USA), used between 350 and 2400~nm, a deuterium lamp (McPherson Mo.634, MA, USA), for relative calibration below 350~nm, and a blackbody source (Landcal R1500T, Ametek Land, UK) for the MIR range.
The SNR of the detectors was characterized by statistical sampling of different calibration images.
Uncertainties on the measured spectral radiance were propagated through Eqn.~\ref{eqn:OES_calib}, and included detector noise, calibration source intensity, its spatial uniformity, as well as drift and reciprocity corrections. These contributions were considered uncorrelated and statistically symmetrical. Wavelength calibration was performed against low-pressure $\ce{Hg}$ and $\ce{Ar}$ lamps, while the Instrument LineShape (ILS) and Spatial Resolution Function (SRF) were fitted to combinations of Gaussian and Lorentzian functions to represent the spectral and spatial resolution of the optical system.

\subsubsection{N and O State Density Measurements}

Electronic state densities of atomic species are individually measured from the observed emission features, summarized in Table \ref{tab:lines} and Fig. \ref{fig:energylvls}. 
Assuming optically thin radiance, the normalized upper state population of the observed transition, $n_j/g_j$, is inferred from the integral line intensity according to the following equation
\begin{equation}
	\frac{\int_{\Delta \lambda} L_\lambda}{D} \approxeq \sum_{j, i} \frac{\Delta E_{ji}}{4 \uppi} \einsteincoeff g_j \frac{n_j}{g_j}, 
    \label{eq:OES_nu_over_gu}
\end{equation}
where $L_\lambda$ is the measured line of sight radiance, $D$ is the tube diameter, and $\Delta\lambda$ is a suitable wavelength interval containing the multiplet transitions. A linear baseline subtraction is operated on $L_\lambda$ prior to integration to remove underlying emission contributions from different systems, such as $\ce{N2(B\rightarrow A)}$, as well as any background continuum.

\setlength{\tabcolsep}{5pt}
\begin{table}
\centering
\renewcommand{\arraystretch}{1.15}
\caption{Atomic transitions considered in this work, enabling measurements of three O-atom and six N-atom electronic levels.}\label{tab:lines}\vspace{-3mm}
\small
%\resizebox{\textwidth}{!}{
\begin{tabular}{cccccccccc}
\hline \hline
 \textbf{Species} & \textbf{Wavelength} &\textbf{ Lower level} & \textbf{Upper level} & $\mathbf{g_i}$ & $\mathbf{g_j}$ & $\mathbf{\mathcal{A}_{ji}}$ & $\mathbf{E_i}$ & $\mathbf{E_j}$ & \textbf{TDLAS/OES} \\
 & $\mathrm{\mathring{A}}$ & i & j &  &  & $\mathrm{s^{-1}}$ & $\mathrm{eV}$ & $\mathrm{eV}$ &  \\
 \hline
 \multirow{3}{*}{O} & 7771.944 & $ \specnot{3s}{5}{S}{2}{\circ}  $ & $ \specnot{3p}{5}{P}{3}{} $ & 5 & 7 & 3.69e+07 & 9.146 & 10.741 & \textbf{Yes/Yes} \\ \cline{10-10}
 & 7774.166 & $ \specnot{3s}{5}{S}{2}{\circ}  $ & $ \specnot{3p}{5}{P}{2}{} $ & 5 & 5 & 3.69e+07 & 9.146 & 10.740 & \multirow{2}{*}{No/Yes}\\
 & 7775.388 & $ \specnot{3s}{5}{S}{2}{\circ}  $ & $ \specnot{3p}{5}{P}{1}{} $ & 5 & 3 & 3.69e+07 & 9.146 & 10.740 &  \\
 \hline
 \multirow{3}{*}{O} & 8446.247 & $ \specnot{3s}{5}{S}{1}{\circ} $ & $ \specnot{3p}{3}{P}{0}{} $ & 3 & 1 & 3.22e+07 & 9.521 & 10.989 & \multirow{3}{*}{No/Yes} \\
 & 8446.359 & $ \specnot{3s}{5}{S}{1}{\circ} $ & $ \specnot{3p}{3}{P}{2}{} $ & 3 & 5 & 3.22e+07 & 9.521 & 10.989 & \\
 & 8446.758 & $ \specnot{3s}{5}{S}{1}{\circ} $ & $ \specnot{3p}{3}{P}{1}{} $ & 3 & 3 & 3.22e+07 & 9.521 & 10.989 &  \\
 \hline
 \multirow{3}{*}{O} & 9265.826 & $ \specnot{3p}{5}{P}{3}{} $ & $ \specnot{3d}{5}{D}{2}{\circ} $ & 7 & 5 & 2.97e+06 & 10.741 & 12.079 & \multirow{3}{*}{Yes/No} \\
& 9265.932 & $ \specnot{3p}{5}{P}{3}{} $ & $ \specnot{3d}{5}{D}{3}{\circ} $ & 7 & 7 & 1.48e+07 & 10.741 & 12.079 &  \\
& 9266.006 & $ \specnot{3p}{5}{P}{3}{} $ & $ \specnot{3d}{5}{D}{4}{\circ} $ & 7 & 9 & 4.45e+07 & 10.741 & 12.079 &  \\
 \hline
 \multirow{6}{*}{O} & 11286.317 & $ \specnot{3p}{3}{P}{1}{} $ & $ \specnot{3d}{3}{D}{2}{\circ} $ & 3 & 5 & 2.32e+07 & 10.989 & 12.087 & \multirow{6}{*}{Yes/No} \\
& 11286.406 & $ \specnot{3p}{3}{P}{1}{} $ & $ \specnot{3d}{3}{D}{1}{\circ} $ & 3 & 3 & 1.29e+07 & 10.989 & 12.087 &  \\
& 11286.914 & $ \specnot{3p}{3}{P}{2}{} $ & $ \specnot{3d}{3}{D}{3}{\circ} $ & 5 & 7 & 3.09e+07 & 10.989 & 12.087 &  \\
& 11287.029 & $ \specnot{3p}{3}{P}{2}{} $ & $ \specnot{3d}{3}{D}{2}{\circ} $ & 5 & 5 & 7.74e+06 & 10.989 & 12.087 &  \\
& 11287.118 & $ \specnot{3p}{3}{P}{2}{} $ & $ \specnot{3d}{3}{D}{1}{\circ} $ & 5 & 3 & 8.60e+05 & 10.989 & 12.087 &  \\
& 11287.318 & $ \specnot{3p}{3}{P}{0}{} $ & $ \specnot{3d}{3}{D}{1}{\circ} $ & 1 & 3 & 1.72e+05 & 10.989 & 12.087 &  \\
\hline \hline
\multirow{3}{*}{N} & 7423.641 & $ \specnot{3s}{4}{P}{1/2}{} $ & $ \specnot{3p}{4}{S}{3/2}{\circ} $ & 2 & 4 & 5.64e+06 & 10.326 & 11.996 &  \multirow{3}{*}{No/Yes} \\
 & 7442.298 & $ \specnot{3s}{4}{P}{3/2}{} $ & $ \specnot{3p}{4}{S}{3/2}{\circ} $ & 4 & 4 & 1.19e+07 & 10.330 & 11.996 &  \\
 & 7468.312 & $ \specnot{3s}{4}{P}{5/2}{} $ & $ \specnot{3p}{4}{S}{3/2}{\circ} $ & 6 & 4 & 1.96e+07 & 10.336 & 11.996 &  \\
 \hline
\multirow{7}{*}{N} & 8184.861 & $ \specnot{3s}{4}{P}{3/2}{} $ & $ \specnot{3p}{4}{P}{5/2}{\circ} $ & 4 & 6 & 8.21e+06 & 10.330 & 11.844 &  \multirow{7}{*}{No/Yes} \\
 & 8188.012 & $ \specnot{3s}{4}{P}{1/2}{} $ & $ \specnot{3p}{4}{P}{3/2}{\circ} $ & 2 & 4 & 1.25e+07 & 10.326 & 11.840 & \\
 & 8200.357 & $ \specnot{3s}{4}{P}{1/2}{} $ & $ \specnot{3p}{4}{P}{1/2}{\circ} $ & 2 & 2 & 4.68e+06 & 10.326 & 11.837 &  \\
 & 8210.715 & $ \specnot{3s}{4}{P}{3/2}{} $ & $ \specnot{3p}{4}{P}{3/2}{\circ} $ & 4 & 4 & 5.23e+06 & 10.330 & 11.840 &  \\
 & 8216.336 & $ \specnot{3s}{4}{P}{5/2}{} $ & $ \specnot{3p}{4}{P}{5/2}{\circ} $ & 6 & 6 & 2.26e+07 & 10.336 & 11.844 &  \\
 & 8223.128 & $ \specnot{3s}{4}{P}{3/2}{} $ & $ \specnot{3p}{4}{P}{1/2}{\circ} $ & 4 & 2 & 2.62e+07 & 10.330 & 11.837 &  \\
 & 8242.389 & $ \specnot{3s}{4}{P}{5/2}{} $ & $ \specnot{3p}{4}{P}{3/2}{\circ} $ & 6 & 4 & 1.31e+07 & 10.336 & 11.840 &  \\
 \hline
 \multirow{4}{*}{N} & 8567.735 & $ \specnot{3s}{2}{P}{1/2}{} $ & $ \specnot{3p}{2}{P}{3/2}{\circ} $ & 2 & 4 & 4.86e+06 & 10.680 & 12.126 &  \multirow{4}{*}{No/Yes} \\
 & 8594.000 & $ \specnot{3s}{2}{P}{1/2}{} $ & $ \specnot{3p}{2}{P}{1/2}{\circ} $ & 2 & 2 & 2.09e+07 & 10.680 & 12.122 &  \\
 & 8629.235 & $ \specnot{3s}{2}{P}{3/2}{} $ & $ \specnot{3p}{2}{P}{3/2}{\circ} $ & 4 & 4 & 2.67e+07 & 10.690 & 12.126 &  \\
 & 8655.878 & $ \specnot{3s}{2}{P}{3/2}{} $ & $ \specnot{3p}{2}{P}{1/2}{\circ} $ & 4 & 2 & 1.07e+07 & 10.690 & 12.122 & \\
 \hline
\multirow{8}{*}{N} & 8680.282 & $ \specnot{3s}{4}{P}{5/2}{} $ & $ \specnot{3p}{4}{D}{7/2}{\circ} $ & 6 & 8 & 2.53e+07 & 10.336 & 11.764 & \textbf{Yes/Yes} \\ \cline{10-10}
 & 8683.403 & $ \specnot{3s}{4}{P}{3/2}{} $ & $ \specnot{3p}{4}{D}{5/2}{\circ} $ & 4 & 6 & 1.88e+07 & 10.330 & 11.758 &  \multirow{7}{*}{No/Yes} \\
 & 8686.149 & $ \specnot{3s}{4}{P}{1/2}{} $ & $ \specnot{3p}{4}{D}{3/2}{\circ} $ & 2 & 4 & 1.15e+07 & 10.326 & 11.753 &  \\
 & 8703.247 & $ \specnot{3s}{4}{P}{1/2}{} $ & $ \specnot{3p}{4}{D}{1/2}{\circ} $ & 2 & 2 & 2.16e+07 & 10.326 & 11.750 &  \\
 & 8711.703 & $ \specnot{3s}{4}{P}{3/2}{} $ & $ \specnot{3p}{4}{D}{3/2}{\circ} $ & 4 & 4 & 1.29e+07 & 10.330 & 11.753 &  \\
 & 8718.837 & $ \specnot{3s}{4}{P}{5/2}{} $ & $ \specnot{3p}{4}{D}{5/2}{\circ} $ & 6 & 6 & 6.54e+06 & 10.336 & 11.758 &  \\
 & 8728.901 & $ \specnot{3s}{4}{P}{3/2}{} $ & $ \specnot{3p}{4}{D}{1/2}{\circ} $ & 4 & 2 & 3.75e+06 & 10.330 & 11.750 &  \\
 & 8747.369 & $ \specnot{3s}{4}{P}{5/2}{} $ & $ \specnot{3p}{4}{D}{3/2}{\circ} $ & 6 & 4 & 9.65e+05 & 10.336 & 11.753 &  \\
 \hline 
 \multirow{9}{*}{N} & 10105.132 & $ \specnot{3p}{4}{D}{1/2}{\circ} $ & $ \specnot{3d}{4}{F}{3/2}{} $ & 2 & 4 & 2.81e+07 & 11.750 & 12.977 &  \multirow{9}{*}{No/Yes} \\
 & 10108.892 & $ \specnot{3p}{4}{D}{3/2}{\circ} $ & $ \specnot{3d}{4}{F}{5/2}{} $ & 4 & 6 & 3.02e+07 & 11.753 & 12.979 &   \\
 & 10112.481 & $ \specnot{3p}{4}{D}{5/2}{\circ} $ & $ \specnot{3d}{4}{F}{7/2}{} $ & 6 & 8 & 3.41e+07 & 11.758 & 12.983 &   \\
 & 10114.640 & $ \specnot{3p}{4}{D}{7/2}{\circ} $ & $ \specnot{3d}{4}{F}{9/2}{} $ & 8 & 10 & 3.90e+07 & 11.764 & 12.989 &   \\
 & 10128.279 & $ \specnot{3p}{4}{D}{3/2}{\circ} $ & $ \specnot{3d}{4}{F}{3/2}{} $ & 4 & 4 & 9.84e+06 & 11.753 & 12.977 &   \\
 & 10147.267 & $ \specnot{3p}{4}{D}{5/2}{\circ} $ & $ \specnot{3d}{4}{F}{5/2}{} $ & 6 & 6 & 7.32e+06 & 11.758 & 12.979 &   \\
 & 10164.848 & $ \specnot{3p}{4}{D}{7/2}{\circ} $ & $ \specnot{3d}{4}{F}{7/2}{} $ & 8 & 8 & 3.83e+06 & 11.764 & 12.983 &   \\
 & 10166.801 & $ \specnot{3p}{4}{D}{5/2}{\circ} $ & $ \specnot{3d}{4}{F}{3/2}{} $ & 6 & 4 & 5.64e+05 & 11.758 & 12.977 &  \\
 & 10199.995 & $ \specnot{3p}{4}{D}{7/2}{\circ} $ & $ \specnot{3d}{4}{F}{5/2}{} $ & 8 & 6 & 2.39e+05 & 11.764 & 12.979 &   \\
 \hline \hline
\end{tabular}
\end{table}

OES and TDLAS measurements can be further combined to relax underlying assumptions when inferring internal state populations.
As mentioned above, the analytical formulation presented in Eqn. \ref{eq:4} assumed a Boltzmann ratio between the lower and upper states of the probed electronic transition at the measured translational temperature. This was useful to estimate the upper state density, required to account for the stimulated emission, thus yielding a relationship between the integral absorbance and the lower state density only.
Additionally, Eqn. \ref{eq:OES_nu_over_gu} assumed optically thin radiance to infer the upper state population from the measured emission, while strong atomic lines are expected to be partially self-absorbed.

To this purpose, we consider the radiative transfer equation for a homogeneous slab crossing the shock tube, leading to the following system of equations,
\begin{equation}
	\begin{cases}
		\alpha_{\lambda} =  - \ln \left(\frac{I\subrm{t}}{I\subrm{0}}\right) = \kappa_{\lambda} \times D \\
		L_{\lambda} = \frac{\varepsilon\subrm{\lambda}}{\kappa\subrm{\lambda}} \times [1- \exp(-\kappa_{\lambda} D)] \approx \varepsilon\subrm{\lambda} D \times (1- \frac{\kappa_{\lambda} D}{2}) \\
		\kappa_{\lambda} = \sum_{j, i} \frac{\lambda_{ji}^4}{8 \uppi \speedoflight} g_j \einsteincoeff \left(\frac{n_i}{g_i} - \frac{n_j}{g_j}\right) \phi_{ij}  = \sum_{j, i} K_{ji} \left(\frac{n_i}{g_i} - \frac{n_j}{g_j}\right) \phi_{ij}   \\
		\varepsilon\subrm{\lambda} = \sum_{j, i} \frac{\Delta E_{ji}}{4\uppi} g_j \einsteincoeff \frac{n_j}{g_j} \phi_{ij} = \sum_{j, i} C_{ji} \frac{n_j}{g_j} \phi_{ij} \\
		
	\end{cases}\label{eq:OES+TDLAS}
\end{equation}
The first equation recasts Eqn. \ref{eq:4} in terms of wavelength, while the right-hand side of the second equation is a second-order approximation of the homogeneous radiative transfer equation (RTE). In cases where self absorption is particularly strong, higher order terms can also be considered.
The last two equations also define the absorption and emission linestrength coefficients, $K_{ji}$ and $C_{ji}$, respectively.

Noting that the upper- and lower-state energies within the multiplet are sufficiently close, the quantities $n_i/g_i$ and $n_j/g_j$ may be reasonably assumed to remain constant over the indices $i$ and $j$, respectively, assuming their energy separation is small such that each state has nearly the same degeneracy-weighted Boltzmann population (Eq. \ref{eq:BoltzPop}).

\begin{equation}
	\frac{n_j}{g_j} = n_{total}^s \frac{\exp\big({-\frac{E_j}{k_BT}}\big)}{Q}
    \label{eq:BoltzPop}
\end{equation}

In addition, assuming that all transitions exhibit the same lineshape, the system admits the following analytical solution.

\begin{equation}
	\begin{cases}
		\frac{n_i}{g_i} - \frac{n_j}{g_j} = \frac{\int \alpha_{\lambda} / D }{\sum K_{ji}} \\
		%\frac{n_j}{g_j} = \frac{\int L_{\lambda} /D}{\sum C_{ji} - \left(\frac{n_i}{g_i} - \frac{n_j}{g_j}\right) \sum C_{ji} \sum K_{ji} \frac{D}{2} \int \phi_{k}^{2}(\lambda_{ij}) }.
        \frac{n_j}{g_j} = \frac{\int L_{\lambda} /D}{\sum C_{ji} - \left(\frac{n_i}{g_i} - \frac{n_j}{g_j}\right)  \frac{D}{2} \sum \sum C_{ji} K_{mn}  \langle \phi_{ij}, \phi_{mn} \rangle }.
	\end{cases}
\end{equation}
The inner product can be simplified in the limiting cases of non-overlapping transitions, such as those considered here, giving 
\begin{equation}
\frac{n_j}{g_j} =
    \frac{\int L_{\lambda} /D}{\sum C_{ji} - \left(\frac{n_i}{g_i} - \frac{n_j}{g_j}\right)  \frac{D}{2} \sum C_{ji} K_{ji}  \int \phi^2(\lambda) } 
\end{equation}
Self absorption leads to a term proportional to the square of the lineshape, which is measured by TDLAS with negligible instrumental broadening, thus allowing the evaluation of the integral.
In contrast, the OES instrumental broadening is non-negligible, and requires a large enough wavelength range to include the energy smeared by the instrument lineshape to evaluate  $\int L_{\lambda}$.
This combination of OES and TDLAS thus enables improved inferences of absorbing and emitting states for transition wavelengths where both were measured and isolated from overlapping emission lines (i.e., for the 777 nm and 868 nm transitions).

In addition to measurements of lower and upper states, the combination of TDLAS and OES enabled nine different energy levels to be probed -- three for O and six for N. With so many electronic states resolved, a Boltzmann temperature can be calculated from the measured upper and lower state densities via Eqn. \ref{BoltzTemp}. Pairwise electronic temperatures are calculated directly with Eqn. \ref{BoltzTemp}, and a least-squares fit is used to calculate $T_{\mathrm{elec}}$ when more than two states are used -- i.e. Eqn. \ref{eq:ElecLSq} where n$_i$/g$_i$ is modeled via Eqn. \ref{eq:BoltzPop} while allowing $T_{\mathrm{elec}}$ and n$_{total}^s$ as fit parameters.

\begin{equation}
\label{BoltzTemp}
	T_{elec} = -\frac{\Delta E_{ji}} {\boltzmannconstant \ln (\frac{n_j/g_j}{n_i/ g_i} )}.
\end{equation}

\begingroup
\small
\begin{equation}\label{eq:ElecLSq}
   \arg\min\left(\sum_m\left(\begin{bmatrix}
    \ln(n_1/g_1)_{\mathrm{model}}\\
    \vdots\\
    \ln(n_m/g_m)_{\mathrm{model}}\\
\end{bmatrix} - 
\begin{bmatrix}
    \ln(n_1/g_1)_{\mathrm{meas}}\\
    \vdots\\
    \ln(n_m/g_m)_{\mathrm{meas}}\\
\end{bmatrix}\right)^2 \right)
\end{equation}
\endgroup

TDLAS measurements alone could be used to infer a $T_{\mathrm{elec}}$ for O-atom states, but inferences of $T_{\mathrm{elec}}$  for N-atom states would not have been possible without the inclusion of upper states probed via emission.
For all experiments, the combination of TDLAS and OES measurements allows a comparison between the two methods as well as demonstrating how they complement one another.

%=======================================================================================================
\section{Kinetics Models}

Experiments leveraged both TDLAS and OES for inferences of NO modal temperatures and number densities for V$_{shock}$ < 5 km/s and O and N state population number densities and electronic temperatures for V$_{shock}$ > 5 km/s, providing valuable validation data for comparison to kinetics models.
Thus, three types of models were pursued -- two one-dimensional, incident-shock-based codes and as well as a two-dimensional, axisymmetric sphere simulation with the NASA Data Parallel Line Relaxation Code (DPLR) simulation framework.
These three models provide comparisons for both the emission- and absorption-inferred temperature and number density results, although suggested changes to these models falls outside the scope of this work.
The current work prioritizes demonstrating the combined diagnostic strategy at selected velocity conditions, although future evaluation of the full velocity range can enable refinement of key reaction rates.
A brief description of each is provided to emphasize points of distinction for discussion in later sections.

The two 1-D models differ from the DPLR model due to their respective treatment of the flowfield.
Treatment in the DPLR model assumes similarity between the centerline flow in EAST and the stagnation line of a sphere \cite{dplr,brandis2016nonequi}.
For this similarity assumption, a sphere with 3 m radius is typically used to create a large shock standoff distance.
This large standoff distance better recreates the shock tube test slug \cite{Fagnani2026,Fagnani2025}.
The resulting DPLR simulation provides spatial post-shock profiles that can be transformed for the spatial and temporal shock tube results \cite{Chang2023,Drescher2026}.
Diffusional broadening of the shock front is included.
Reaction rates are taken in part from the Park93 model and in part from Cruden and Brandis, which provides corrected rates involving the Zeldovich reactions and some dissociation reactions \cite{park,dplr}.

In the 1-D models, conservations of mass, momentum, and energy are solved subject to the evolution of vibrational relaxation and chemical reactions, with the two 1-D models differing only by the assumed kinetic model.
Both 1-D solvers use a jump condition at the shock interface and track the subsequent endothermic reactions, with the initial post-incident-shock temperature and pressure calculated from the incident shock velocity using the Stanford frozen-chemistry shock calculator, FROSH \cite{frosh} informed with NASA thermochemical polynomial fits \cite{McBride2002} to enable calculation of the frozen and vibrationally equilibrated conditions for use implementing chemical kinetic models.
The use of NASA thermochemical polynomial fits \cite{McBride2002} throughout the FROSH and one-dimensional shock simulations ensures all three models leverage the same thermochemistry foundation.
For current modeling, the initial post-shock temperature, pressure, and gas velocity (in a shock-fixed reference frame) is calculated subject to frozen vibrational energy and chemistry.
Subsequently, the post-shock profile is calculated with a coupled time- and space-stepping simulation.
For each time step (typically 1-10 ns of particle time), the vibrational energy and species number densities progress subject to known differential equations \cite{jesse_n2o,jesse_noar,jesse_non2,jesse_o2n2,jesse_vibdiss}.
This calculation provides a new internal energy and composition condition to use for conserving mass, momentum, and energy, resulting in a new temperature, pressure, and gas velocity for the next timestep, with the evolving gas velocity being used to relate the particle time to the distance post-shock.
The result of this calculation is a post-shock spatial distribution of modal temperatures and species number densities, that can be transformed to the laboratory time-frame via the shock velocity.

Although both 1-D models use the same reference frame treatment, they differ based on the reaction rates included.
In the following sections, the Streicher \textit{et al.} model refers to a set of rates sourced from recent experimental studies where available \cite{jesse_n2o,jesse_noar,jesse_non2,jesse_o2n2,jesse_vibdiss,StreicherVibAir_PartI,StreicherVibAir_PartII,StreicherVibAir_PartIII}, although it contains no consideration for ionized species in this model.
Thus, the Streicher \textit{et al.} model is not used for comparison in the > 5 km/s shocks targeting O and N.
The Park model refers to the set of rates from the Park93 model. The Park model includes rates for ionized species, so it can be used for comparison in the high-velocity shocks \cite{park,peter,yang}.

Again, the inclusion of these three types of modeling results primarily serves as a benchmark for TDLAS and OES data, with model evaluation and refinement reserved for future work.
However, facility effects -- including boundary layer growth and corresponding test-time reduction -- were considered alongside the kinetic modeling \cite{Mirels1964,Mirels1966,Clarke2023,Clarke2026}.
To estimate these effects, we computed the one-dimensional solution in the ideal case (i.e. no mass loss to boundary layers) as well as for the Mirel's mass loss models \cite{Mirels1964,Mirels1966}, as shown in Ref.~\cite{fagnani2026b}.
The resulting temperature and species number density profiles were then compared between the two cases, with negligible difference between the models with and without boundary layer effects -- at least for the post-shock distances and times considered for this work.
Thus, the current modeling is presented without a correction for boundary layer effects, although this would need to be updated to compare with data beyond the current 10-20 $\mu$s window considered here.
Together, Table \ref{tab:ModelComp} summarizes many key differences between the three model frameworks as well as the consistent treatment of reference frame and boundary layers across the three model frameworks.

\begin{table}
\caption{Comparison of the three kinetic model frameworks.}
\label{tab:ModelComp}
\begin{center}
\begin{tabular}{c c c c}
\hline
\hline
& DPLR & Streicher \textit{et al.} & Park \\
\hline
\textbf{Simulation Setup} & 3-m radius sphere & \multicolumn{2}{c}{1-D shock}\\ \hline
\textbf{Calculation Mode} & Steady-state spatial distribution & \multicolumn{2}{c}{Particle timestep and local velocity} \\ \hline
\textbf{Shock-Jump Condition} & Viscous shock calculation & \multicolumn{2}{c}{Frozen shock calculation} \\ \hline
\textbf{Reference Frame} & \multicolumn{3}{c}{Shock-fixed coordinates} \\
 & \multicolumn{3}{c}{Laboratory time calculated via V$_{shock}$} \\
\hline
\textbf{Boundary Layers} & \multicolumn{3}{c}{Neglected \cite{Clarke2026,Clarke2023,Mirels1964,Mirels1966}} \\
\hline
\hline
\textbf{Reaction Rates} & \multicolumn{3}{c}{} \\ \hline
O$_2$ dissociation & \cite{park} & \cite{jesse_vibdiss,StreicherVibAir_PartI,StreicherVibAir_PartIII} & \cite{park} \\ \hline
N$_2$ dissociation & \cite{park} & \cite{park} & \cite{park} \\ \hline
Zeldovich reactions & \cite{dplr} & \cite{StreicherVibAir_PartI,StreicherVibAir_PartII,StreicherVibAir_PartIII,jesse_n2o} & \cite{park} \\ \hline
NO dissociation & \cite{dplr} & \cite{jesse_noar,jesse_non2,StreicherVibAir_PartII,StreicherVibAir_PartIII} & \cite{park} \\ \hline
Excitation/ionization & \cite{park} & N/A & \cite{park} \\ \hline \hline
\end{tabular}
\end{center}
\end{table}

Although it is not strictly a kinetic model result, all experimental data are also compared to equilibrium calculations from the Chemical Equilibrium with Applications program (CEA), which is a software that calculates chemical equilibrium product concentrations from the provided reactants and shock conditions using the NASA Glenn polynomials \cite{cea,McBride2002}.
These equilibrium calculations are for the nominal V$_{shock}$, although the spatiotemporal coupling for incident shocks along with shock trajectory and deceleration effects mean that some late-time data may deviate from this equilibrium calculation if that gas was initially shock-heated at a higher velocity.
Variation of V$_{shock}$ was not considered in this work, but the long-time TDLAS results could enable future studies of trajectory and boundary layer effects in EAST experiments.

%===============================================================================================================
\section{Results}

The results of this study are divided into three broad categories. First, TDLAS scan data of both incident and transmitted intensity (I$_0$ and I$_t$, respectively) are provided for the seven laser wavelengths to demonstrate the low noise of raw TDLAS data as well as the phase shift between the intensity modulation and the wavelength output that limits upscan vs. downscan data quality.
Subsequently, the combined TDLAS and OES results are provided separately for two NO experiments (shots 66/21 and 66/26) probing MIR wavelengths and one O/N experiment (shot 66/41) probing NIR wavelengths.
Within each of these sections, the results demonstrate the progression from spectrally-resolved absorption or emission data to the corresponding inferences of modal temperatures and species number density.
These inferred quantities are compared with the three aforementioned kinetic models and CEA equilibrium values.

\subsection{TDLAS Scan Characteristics}

\begin{figure}[htb!]
\centering
\subfigure[]
{{\includegraphics[width=0.49\textwidth]{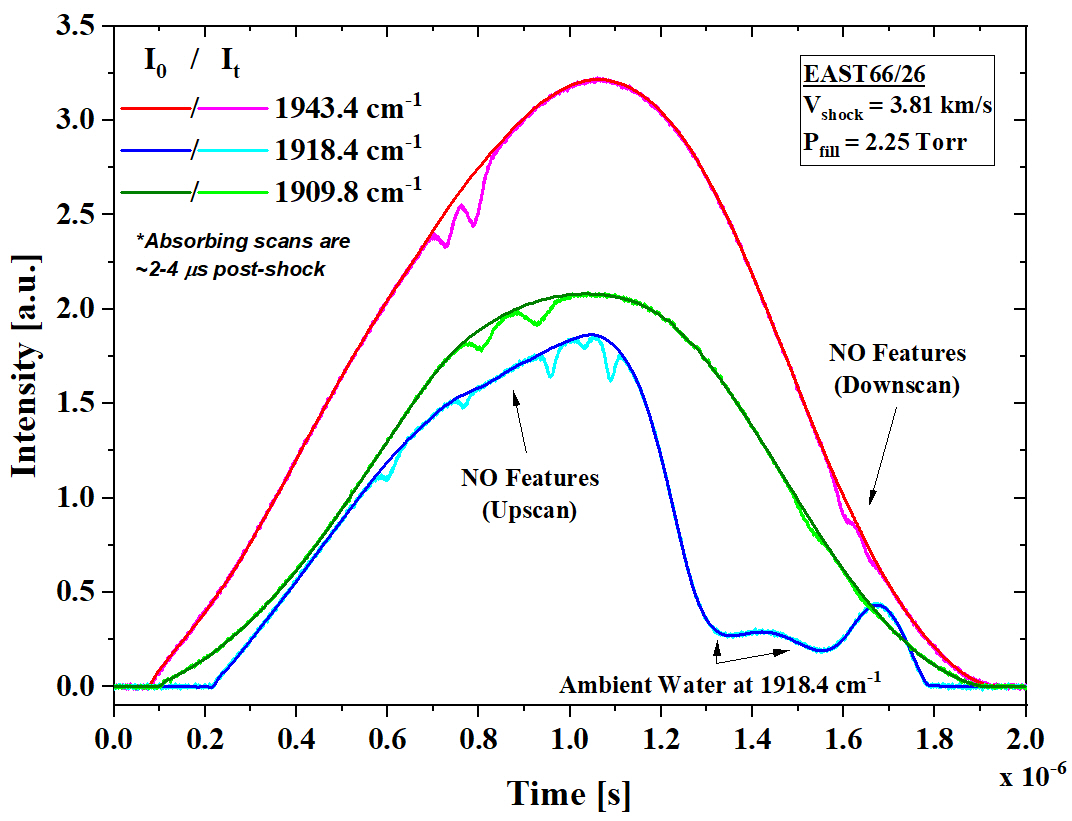}}}%
\subfigure[]
{{\includegraphics[width=0.49\textwidth]{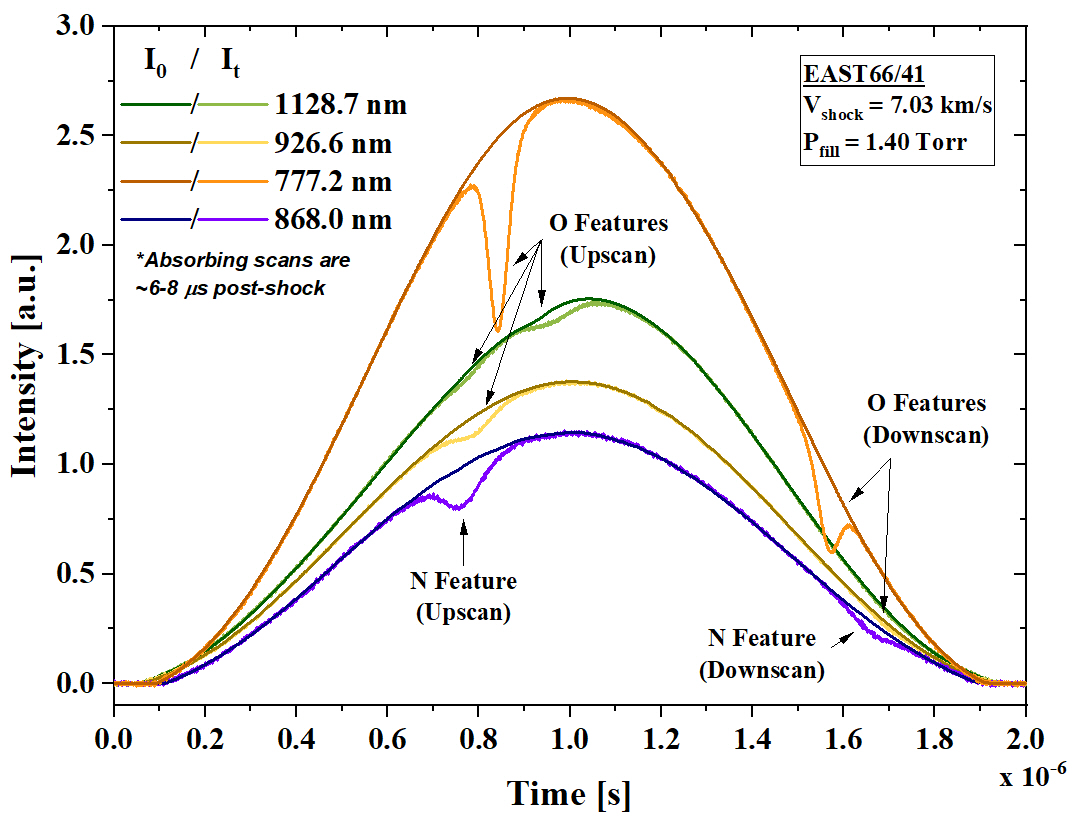}}}%
\caption{\label{fig:raw}Experimental traces of recorded signal intensity for (a) three lasers targeting NO and (b) four lasers targeting O and N states. NO, O, N, and ambient water features are labeled for upscans and downscans.}
\end{figure}

Raw intensity data for both non-absorbing (I$_0$) and absorbing (I$_t$) profiles of the laser scans are compiled in Fig. \ref{fig:raw} for the MIR profiles of NO features and the NIR profiles of O/N features.
In each case, absorption features are clearly observed as dips in the transmitted intensity (I$_t$) relative to the incident profiles (I$_0$), from which absorbance is calculated via Eqn. \ref{eq:beer}.
Most of the laser profiles exhibit strong intensities, helping to minimize the baseline noise of the absorbance data.
Additionally, most profiles are free from interfering absorption, with the exception of 1918 cm$^{-1}$, where a strong ambient water vapor absorption feature is present.
However, this absorbance from water vapor was both stable over time and reasonably separate from the targeted NO features.
These strong, low-noise intensity data provide a larger SNR, aiding the resolution of weaker absorbance (e.g. < 0.1) that may have otherwise been limited by electronic noise for such high-temperature, low-pressure conditions.

Notably, these absorption features also show the phase shift between the laser output intensity modulation and the corresponding wavelength modulation.
If wavelength and intensity were aligned, the absorption features would occur with the same relative symmetry as the intensity profiles.
Instead, the turnaround point for intensity (i.e. the maxima in Fig \ref{fig:raw}) occurs before the wavelength turnaround (i.e. the midpoint between upscan and downscan features), with wavelength lagging by approximately 0.2-0.4 $\upmu$s in time.
In general, this phase shift causes the absorption features on the upscan to coincide with higher incident intensity, whereas the downscan coincides with lower incident intensity.
The lasers with the largest phase shift -- specifically 1918 cm$^{-1}$ and 1128.7 nm -- exhibit a shift large enough to entirely obscure the downscan data.
This provides additional justification for plotting upscan and downscan data separately, as upscan data could leverage all available absorbing states, while downscan data do not include the 1918 cm$^{-1}$ and 1128.7 nm features.
As previously mentioned, this phase shift is inherent in the laser systems.
Although it introduces some time-resolution limitations, the current TDLAS scan profiles still exhibit strong absorption features with sufficiently high time-resolution for resolving nonequilibrium air species in EAST experiments.

%=======================================================================================================
\subsection{Low Velocity NO Rovibrational States}
\begin{figure}[htb!]
\centering
\subfigure[]
{{\includegraphics[width=.49\textwidth]{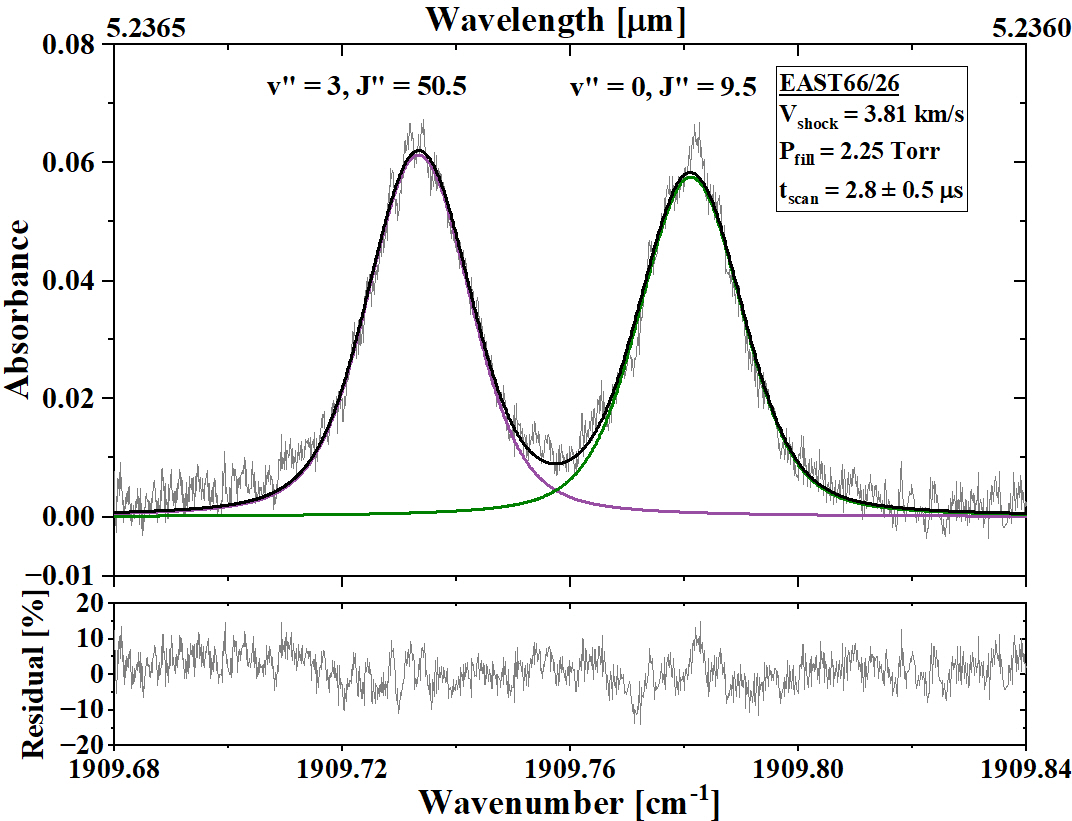}}}%
\subfigure[]
{{\includegraphics[width=.49\textwidth]{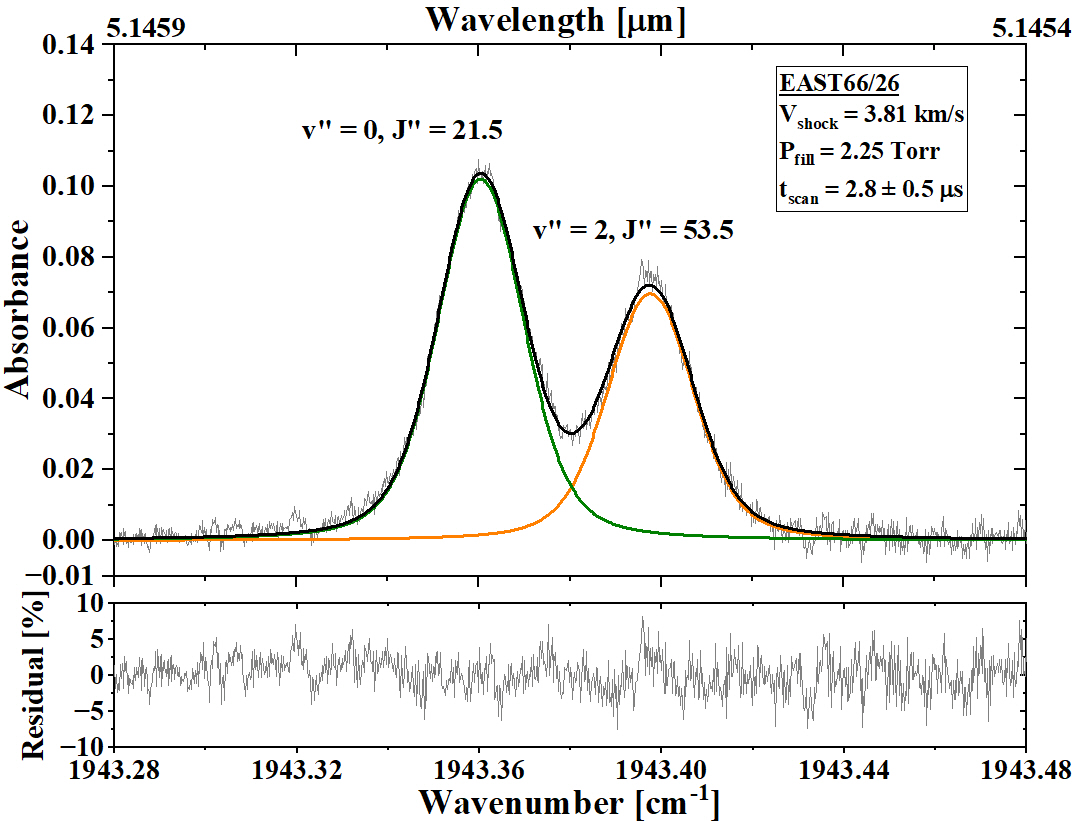}}}\\
\subfigure[]
{{\includegraphics[width=.49\textwidth]{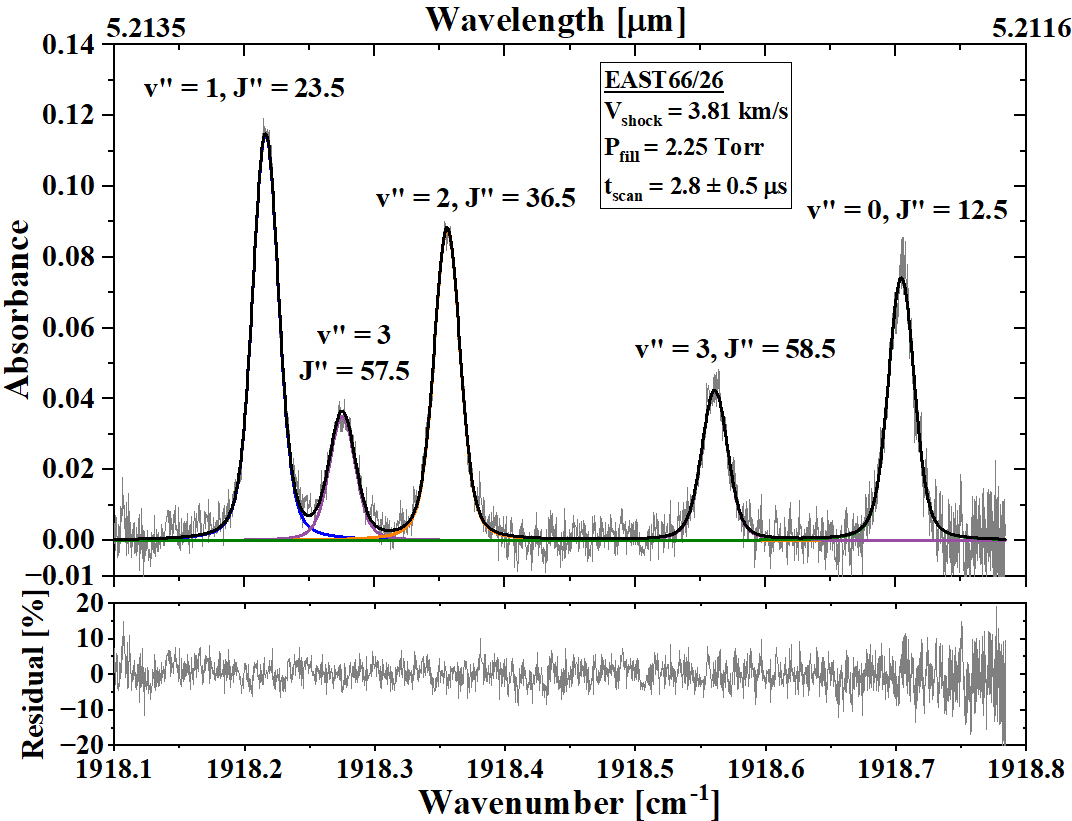}}}%
\caption{\label{fig:fits_individual}Voigt profile fits; (a) 1909.8 cm$^{-1}$; (b) 1943.4 cm$^{-1}$; (c) 1918.4 cm$^{-1}$. Data were recorded for shot 66/26 at approximately 2.8 $\upmu$s. Residuals are presented as a percentage of the maximum absorbance.}
\end{figure}

Two low-velocity conditions -- one high pressure and one low pressure -- are used to compare the TDLAS diagnostic for NO near $ \SI{5.2}{\micro \meter}$ and the OES emission measurements spanning $ \SI{4.9}{\micro \meter}$ - $ \SI{5.5}{\micro \meter}$. Similar NO data were collected across the 3.0-5.0 km/s and 2-10 Torr range of the low-velocity experiments, with shots 66/21 and 66/26 selected to show results for similar shock velocities with differing fill pressures. Fig. \ref{fig:fits_individual} shows example absorbance profiles for scans in the low-pressure condition (shot 66/26), which were calculated from $I_t$ and $I_0$ (Fig. \ref{fig:raw}) using Eqn. \ref{eq:beer}. The resulting absorbance data were then fit with Voigt profiles according to a weighted least-squares scheme, with Voigt fits in Fig. \ref{fig:fits_individual} using the same vibrational state color scheme as Figs. \ref{fig:abs} and \ref{fig:abs_individual}.
From the Voigt fits the integrated absorbance areas could be extracted and the ensemble fit via Eqn. \ref{eq:weightscheme}, leveraging HITEMP spectroscopic constants (Table \ref{tab:NOlines} \cite{hitemp}), across all 9 NO features with varying rotational and vibrational content to obtain $T_{\mathrm{rot}},T_{\mathrm{vib}}, \mathrm{and}\ n_{\mathrm{NO}}$.

\begin{figure}[]
\centering
{{\includegraphics[width=0.49\textwidth]{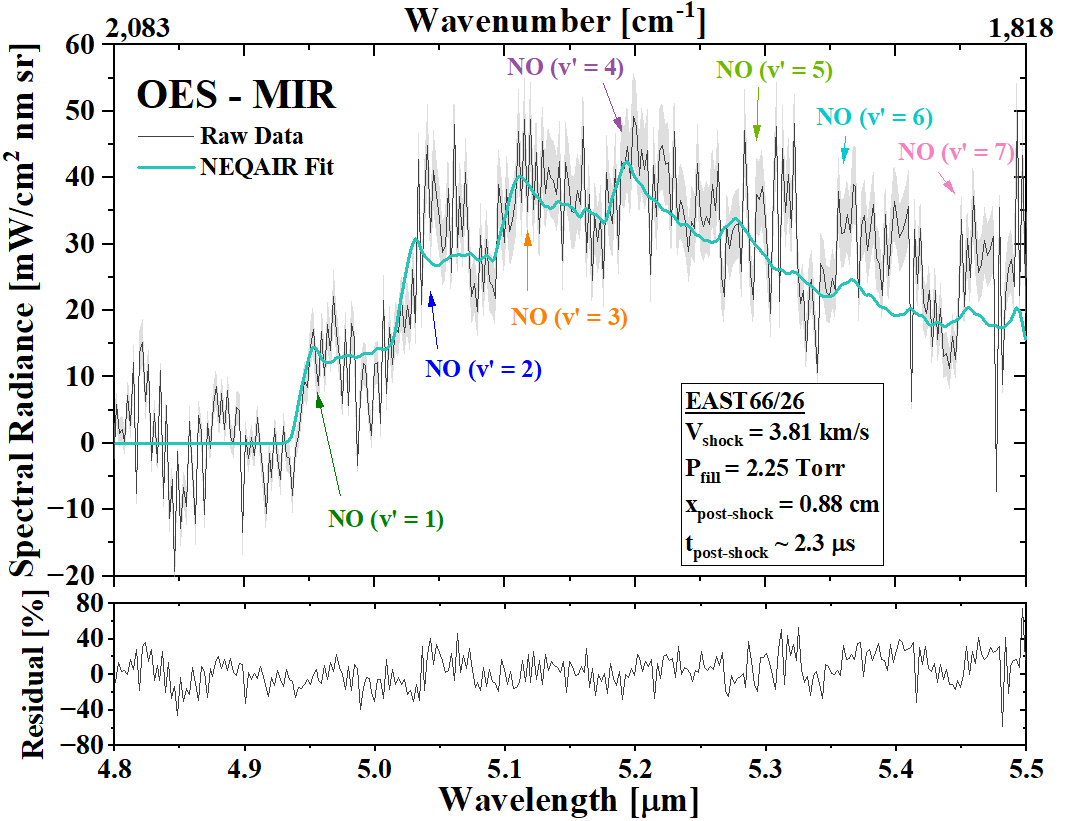}}}%
\caption{\label{fig:66_26_IR_OES_1D_meas_fit_4cm} Spectral radiance and corresponding NEQAIR fit at 8.8~mm from the shock front. Data include shaded uncertainty bounds, and residuals are calculated as a percentage of the maximum NEQAIR-predicted radiance.}
\end{figure}

Fits of the infrared NO emission spectra from OES measurements were also used to extract post-shock density and temperature trends as described in Ref. \cite{cruden2024b} and Thirani \textit{et al.} \cite{thirani}. The spectra are binned in 20-pixel increments, corresponding to a spatial averaging of the radiance over 4.7 mm. This is in addition to an averaging over 7.6 mm (shot 66/26) and 8.0 mm (shot 66/21) due to motion of the shock during the camera exposure time. An example fit is shown in Fig. \ref{fig:66_26_IR_OES_1D_meas_fit_4cm}, corresponding to shot 66/26 at a distance of 8.8 mm behind the shock front, which is approximately the same post-shock timing as the scans in Fig. \ref{fig:fits_individual}. In Fig.  \ref{fig:66_26_IR_OES_1D_meas_fit_4cm}, labels are included to show peaks of the vibrational bands with the same coloring as Fig. \ref{fig:abs} -- although OES probes the upper vibrational state of the transition (i.e. v' = v" + 1). Fluctuation of thermal noise on the camera causes a (presumed) constant offset to the spectrum which is included as a fit parameter.

\begin{figure}[hbt!]
\centering
\subfigure[]
{{\includegraphics[width=0.49\textwidth]{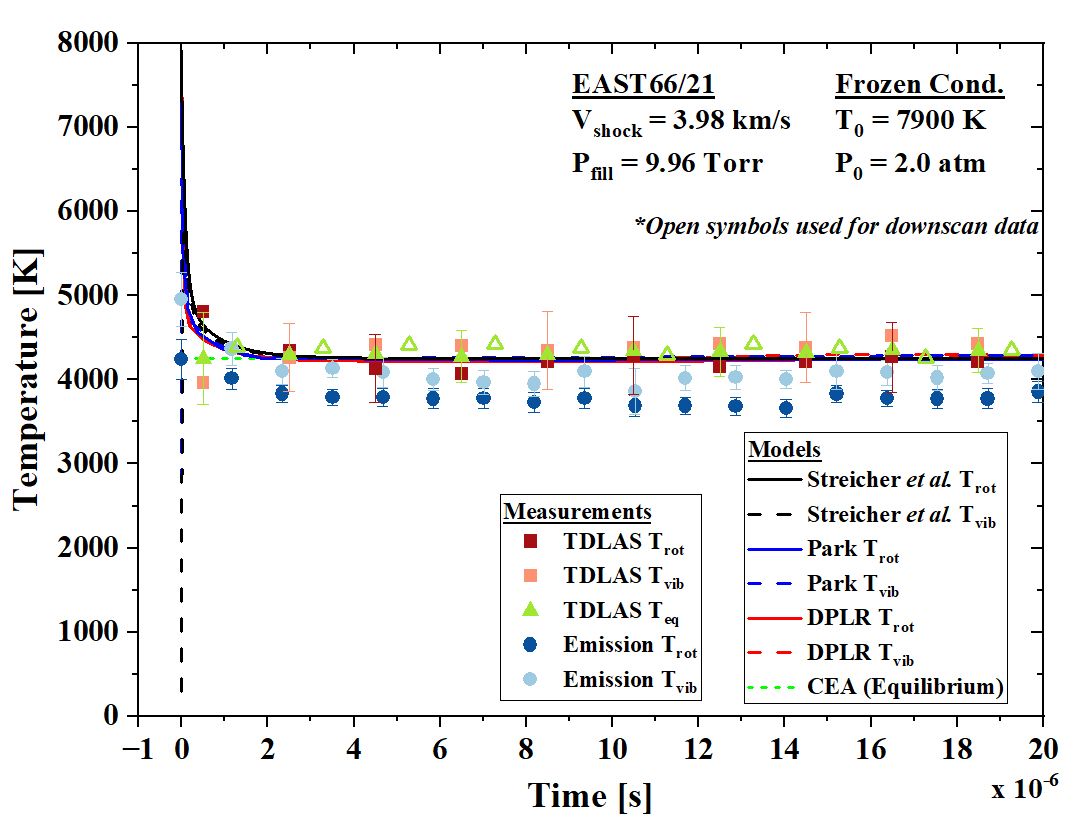}}\label{sfig:TNOeq}}%
\subfigure[]
{{\includegraphics[width=0.49\textwidth]{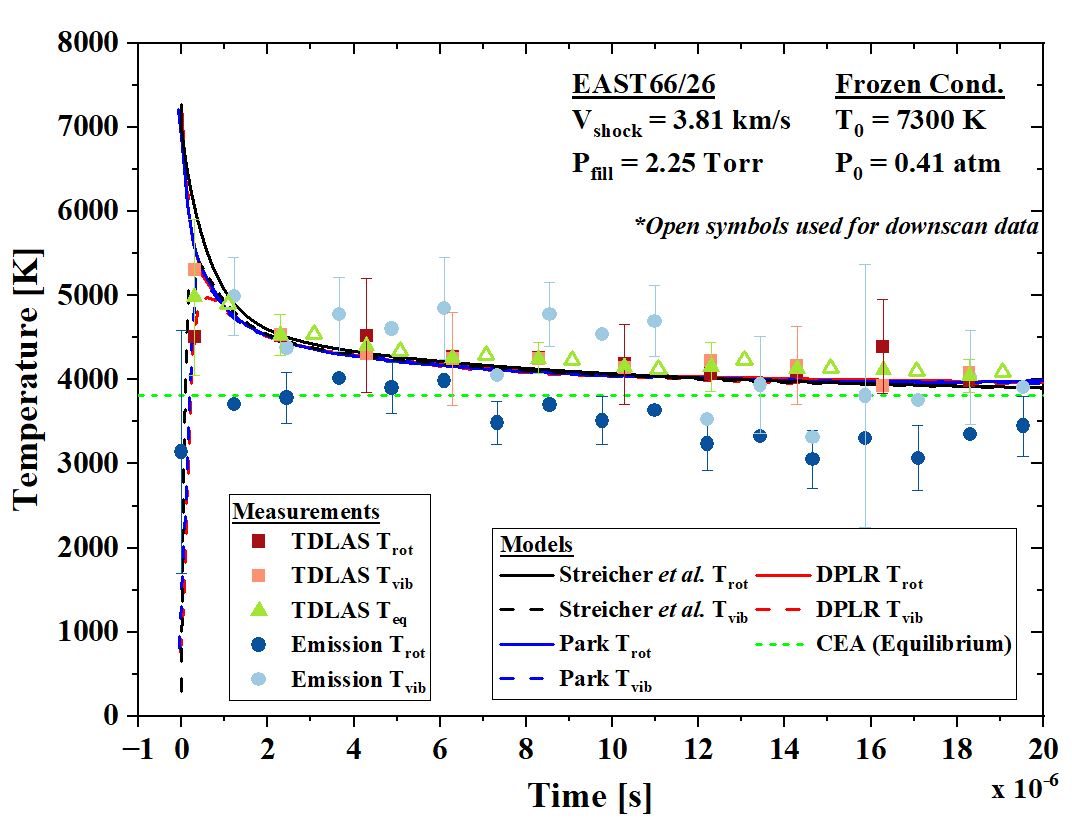}}\label{sfig:TNOneq}}%
\caption{\label{fig:NO_temp}Temperature time-histories for two sample conditions. (a) Equilibrium condition (shot 66/21) -- 3.98 km/s and 9.96 Torr; (b) Nonequilibrium condition (shot 66/26) -- 3.81 km/s and 2.25 Torr.}
\end{figure}

\begin{figure}[hbt!]
\centering
\subfigure[]
{{\includegraphics[width=0.49\textwidth]{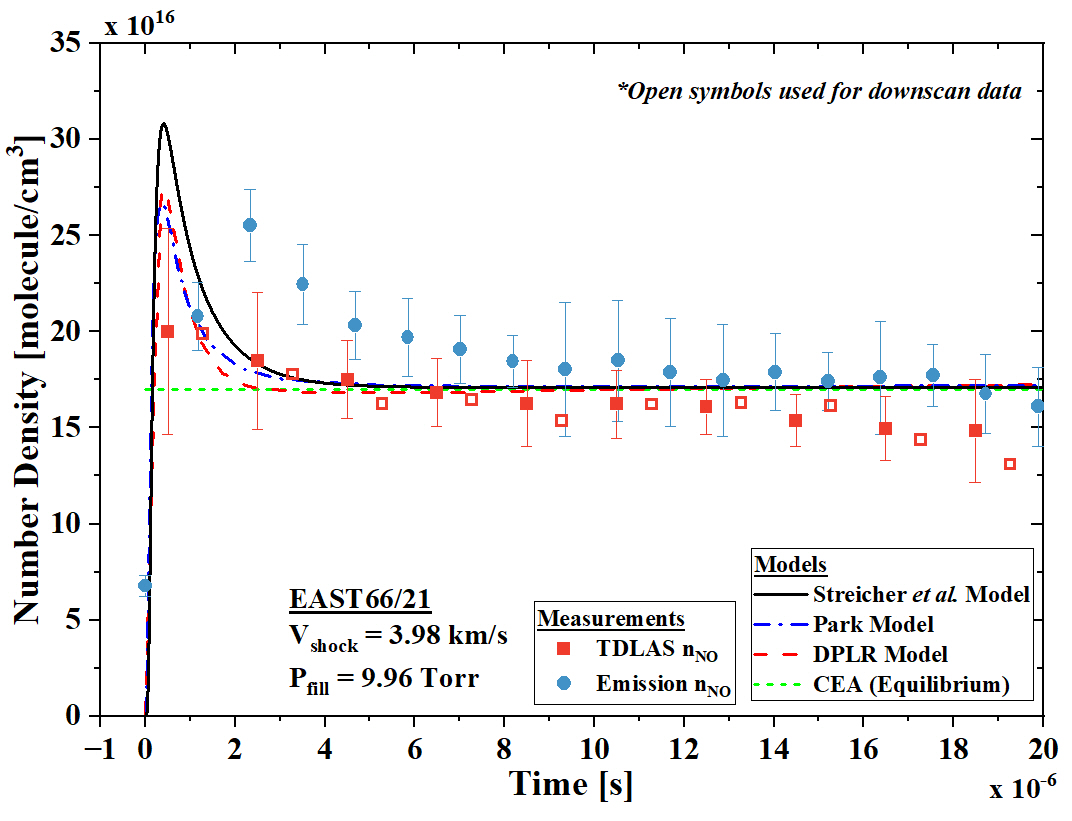}}\label{sfig:nNOeq}}%
\subfigure[]
{{\includegraphics[width=0.49\textwidth]{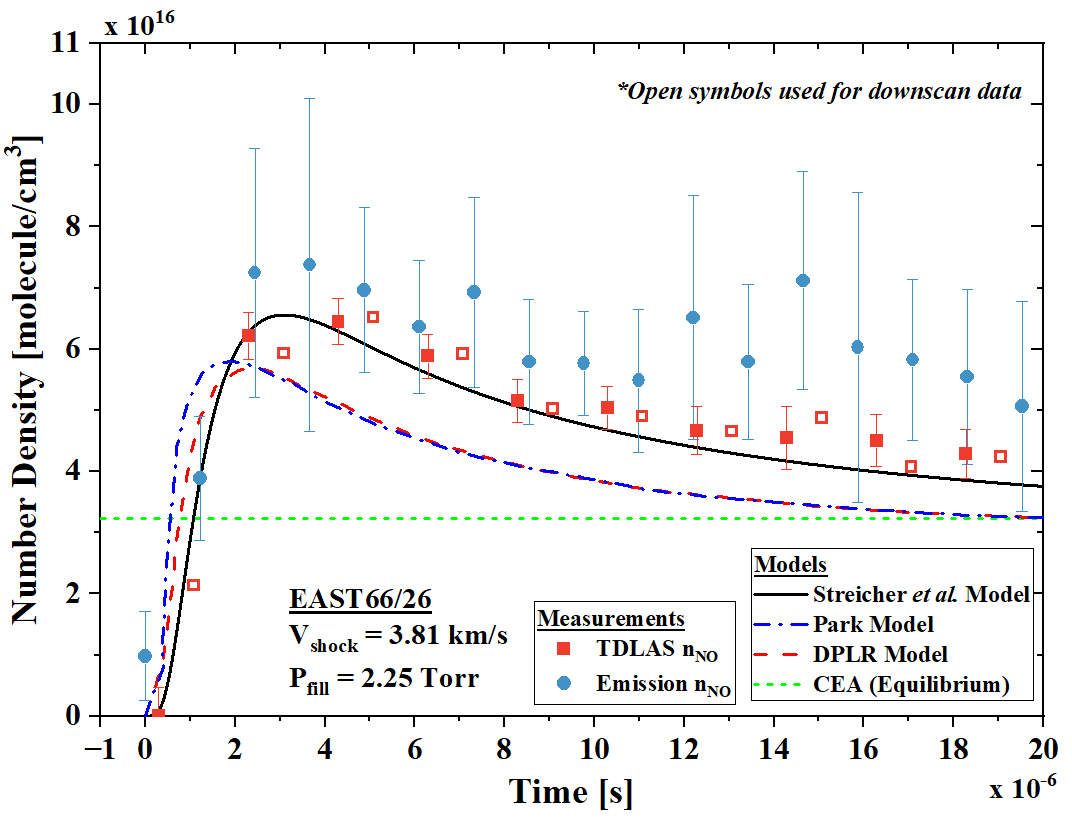}}\label{sfig:nNOneq}}%
\caption{\label{fig:NO_number}NO number density time-histories for two sample conditions. (a) Equilibrium condition (shot 66/21) -- 3.98 km/s and 9.96 Torr; (b) Nonequilibrium condition (shot 66/26) -- 3.81 km/s and 2.25 Torr.}
\end{figure}

The temperature and NO number density time-histories for two lower-velocity conditions are presented in Figs.~\ref{fig:NO_temp} and \ref{fig:NO_number}. The left-hand figures (high-pressure) correspond to shot 66/21 in which thermal and chemical equilibrium is achieved within the test time. The right-hand figures correspond to the lower-pressure shot 66/26 in which non-equilibrium persists.  Emission measurements are included alongside the TDLAS data for comparison, although the need to bin the emission measurements slightly reduces their time-resolution relative to the TDLAS data. 
In each plot, blue markers correspond to emission, while red markers correspond to TDLAS data. Dark colors correspond to $\Trot$ and light colors to $\Tvib$, with open symbols used for downscan data. Green TDLAS data are also included in Fig. \ref{fig:NO_temp} to show a single-temperature fit to all nine states -- in contrast to the two-temperature fits. Finally, three colors of lines are used for the three considered kinetic models, along with the CEA-calculated equilibrium value in light green.

Comparisons of TDLAS and OES measurements remain relatively consistent for both the equilibrium and nonequilibrium cases, with only limited instances where the measurements differ by more than their respective uncertainties.
The strongest deviation between TDLAS and OES occurs for the NO number density results of the equilibrium case (shot 66/21), where OES values substantially exceed TDLAS for the first $ \SI{4}{\micro\second}$ (Fig. \ref{sfig:nNOeq}).
Furthermore, the OES number density results decrease much more slowly than the TDLAS data, although both converged to the equilibrium value at later times (> $ \SI{10}{\micro\second}$).
Likely, the trend in the OES data is strongly influenced by spatial averaging in the emission data that is not present in the TDLAS data.
In contrast, the non-equilibrium condition (shot 66/26) exhibits relatively good consistency between OES and TDLAS throughout the full test time (Fig. \ref{sfig:nNOneq}).
Here, both measurements show similar timescales for NO formation and decay.
OES data trend with slightly higher inferences of NO number density than TDLAS, although the TDLAS results remain within the uncertainty of OES.
In this case, spatial averaging exhibits a much smaller influence on the OES data, as the slower NO formation corresponds to a larger spatial extent of the emission signal.
Finally, OES and TDLAS produce reasonably consistent temperature results for both shots 66/21 and 66/26.
Typically, OES observes a more consistent offset between $T_{\mathrm{rot}}$ and $T_{\mathrm{vib}}$ than is observed from TDLAS, with the single-temperature TDLAS temperature fit exhibiting lower scatter and uncertainty than the two-temperature inferences of either technique.
Given the overlap of probed vibrational states, it is reassuring that the two techniques provide consistent modal temperature and NO number density results, with the remaining discrepancies suggesting the value of employing both TDLAS and OES in EAST experiments.

Comparisons of the TDLAS and OES data with the three kinetic models also demonstrates the value of the combined diagnostic technique.
Generally, all three kinetic models predict similar trends for the modal temperatures, with a rapid post-shock temperature decline and consistent approach to the CEA-predicted equilibrium value (Fig. \ref{fig:NO_temp}).
Slight temperature deviations are present at early times (< $ \SI{1}{\micro\second}$), which can be attributed to differing rates of O$_2$ vibrational relaxation and dissociation among the three models.
These early time deviations quickly converge and remain consistent for the majority of the experimental test time.
Experimental data largely agree with the modeling predictions, although TDLAS data are slightly more consistent with the three kinetic models.
This consistency is especially notable for the nonequilibrium condition (Fig. \ref{sfig:TNOneq}) where the TDLAS observations decrease at a rate very consistent with the three models.
The rapid equilibration for shot 66/21 prevents any comparison of the transient temperature behavior, although TDLAS data again show slightly better consistency with the CEA equilibrium value (Fig. \ref{sfig:TNOeq}).

Models deviate more substantially for the NO number density predictions, as was expected due to the different Zeldovich and NO dissociation rates included in each.
For both conditions, the Streicher \textit{et al.} model predicts a higher peak NO number density and slower NO decay relative to either the Park or DPLR models.
Again, the time-resolution of TDLAS and OES measurements cannot capture rapid equilibration in shot 66/21, but the measurements both capture the expected NO number density at equilibrium (Fig. \ref{sfig:nNOeq}).
At the very end of the test time in shot 66/21, NO number density is observed to decrease slightly, which suggests an influence from shock deceleration \cite{Collen_Decel,Augustin_Decel,Satchell_Decel}.
TDLAS data were collected well beyond $ \SI{20}{\micro\second}$ in each experiment, so the long-time TDLAS data could help refine models of facility effects and boundary layers in future work \cite{Clarke2026,Clarke2023,Mirels1964,Mirels1966,Fagnani2026}.
In the nonequilibrium case (shot 66/26), the deviation of the three models is both substantial and sufficiently long-lived to enable direct comparison with experimental data.
Both TDLAS and OES data show much better consistency with the Streicher \textit{et al.} model, which suggests the slower NO formation and decay rates -- and the correspondingly higher peak NO -- may better predict nonequilibrium air at these conditions.
Although these two conditions provide valuable comparisons for NO number density, similar analysis will be extended to the complete set of shock conditions for Test Series 66.
By analyzing all conditions, the full data set can refine NO reaction rates, especially as they may overlap with reflected shock experiments from which the Streicher \textit{et al.} modeling rates were inferred \cite{StreicherVibAir_PartI,StreicherVibAir_PartII,StreicherVibAir_PartIII}.
Ultimately, the combination of TDLAS and OES measurements in EAST experiments provide valuable new opportunities to refine NO formation and decay for air in the 3-5 km/s regime.

%=======================================================================================================
\subsection{High Velocity N and O Electronic States}
\begin{figure}[htb!]
\centering
\subfigure[]
{{\includegraphics[width=.49\textwidth]{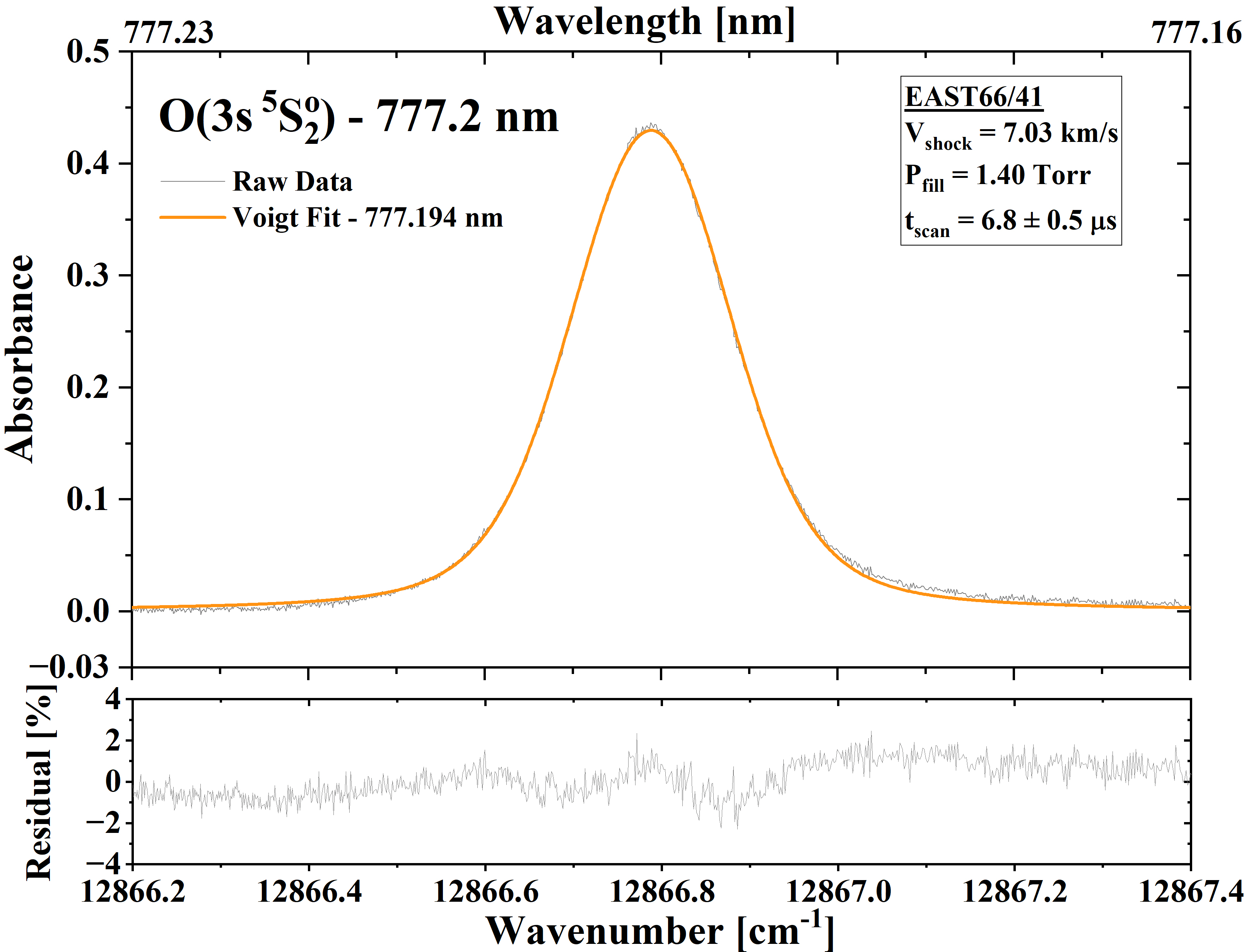}}}%
\subfigure[]
{{\includegraphics[width=.49\textwidth]{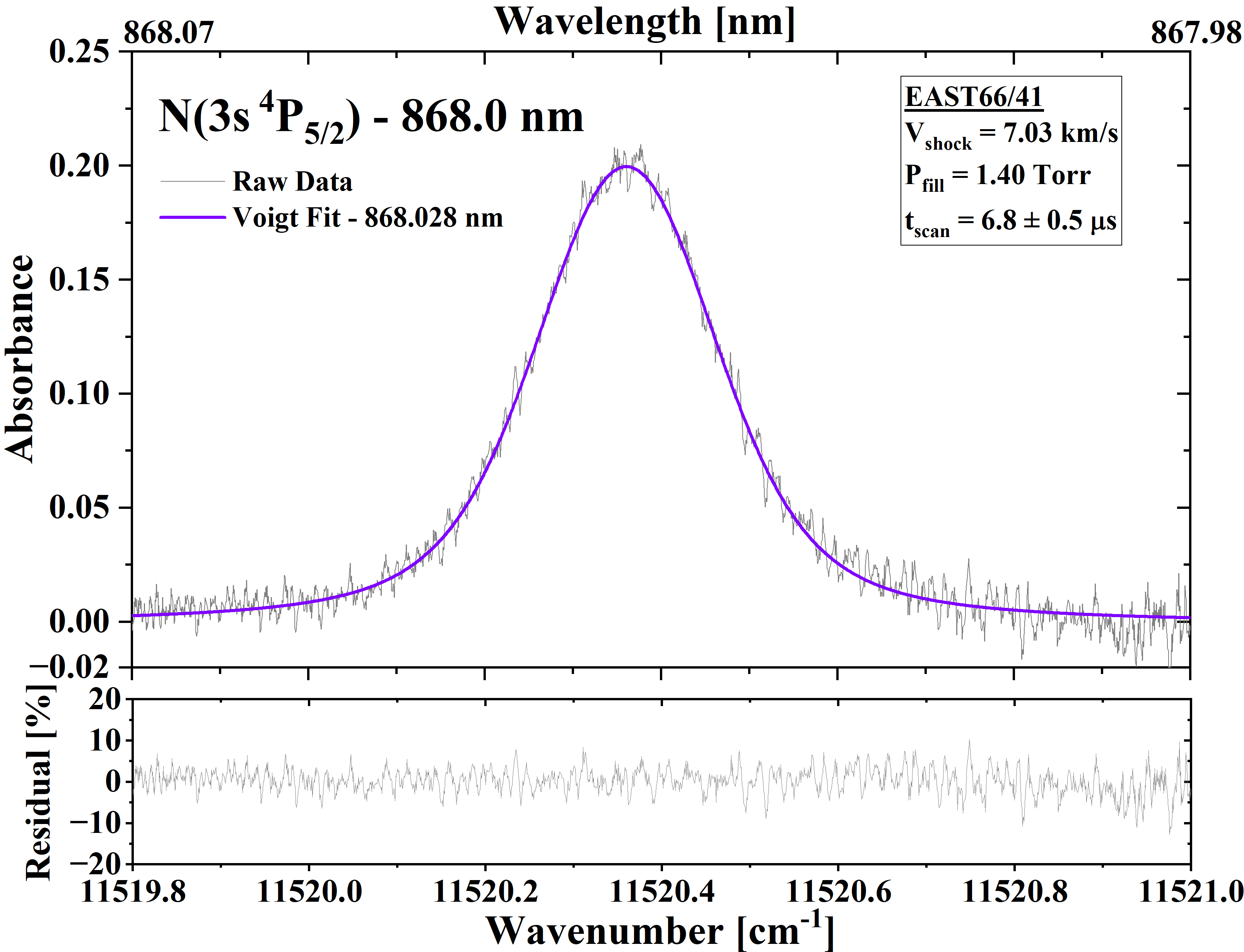}}}\\
\subfigure[]
{{\includegraphics[width=.49\textwidth]{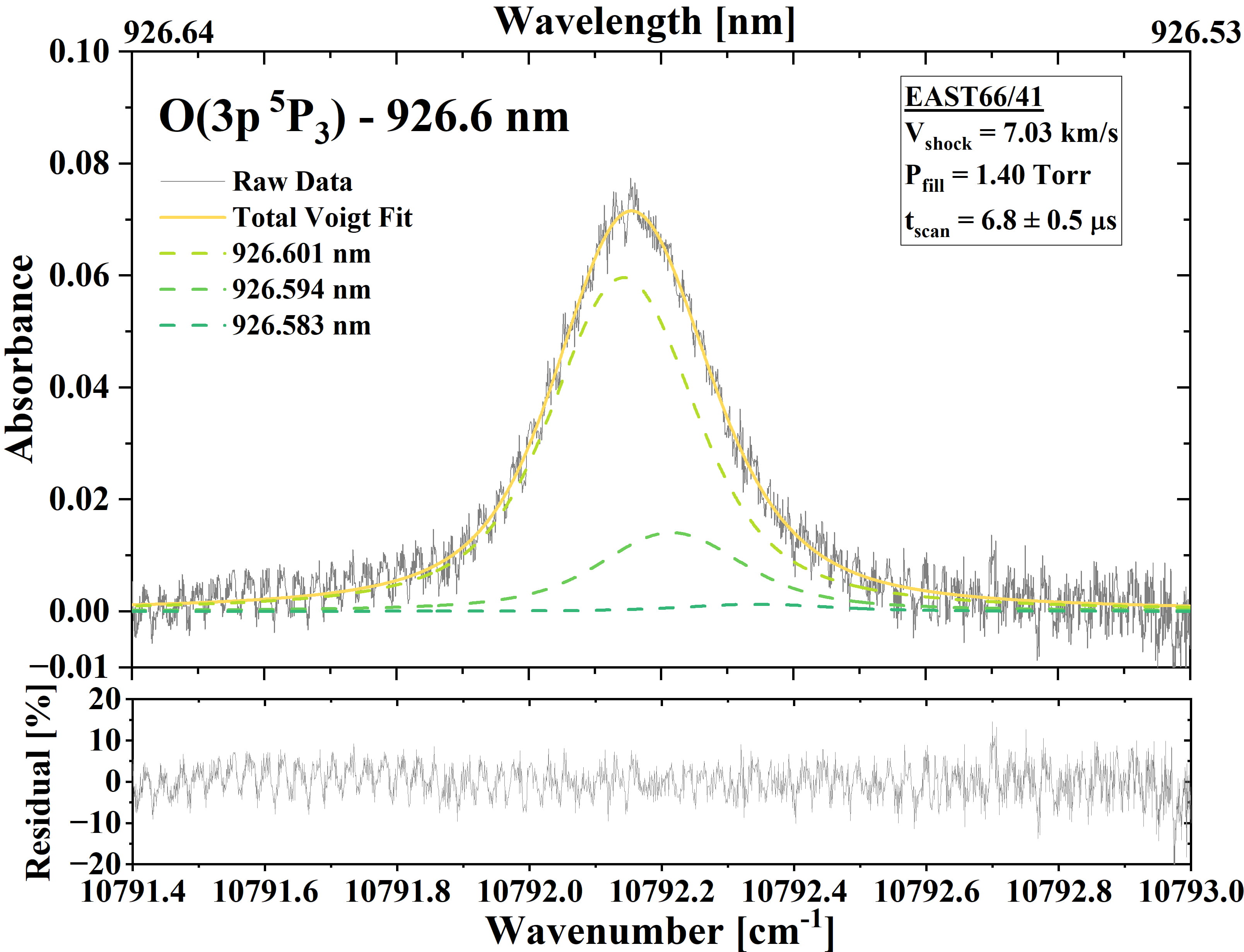}}}%
\subfigure[]
{{\includegraphics[width=.49\textwidth]{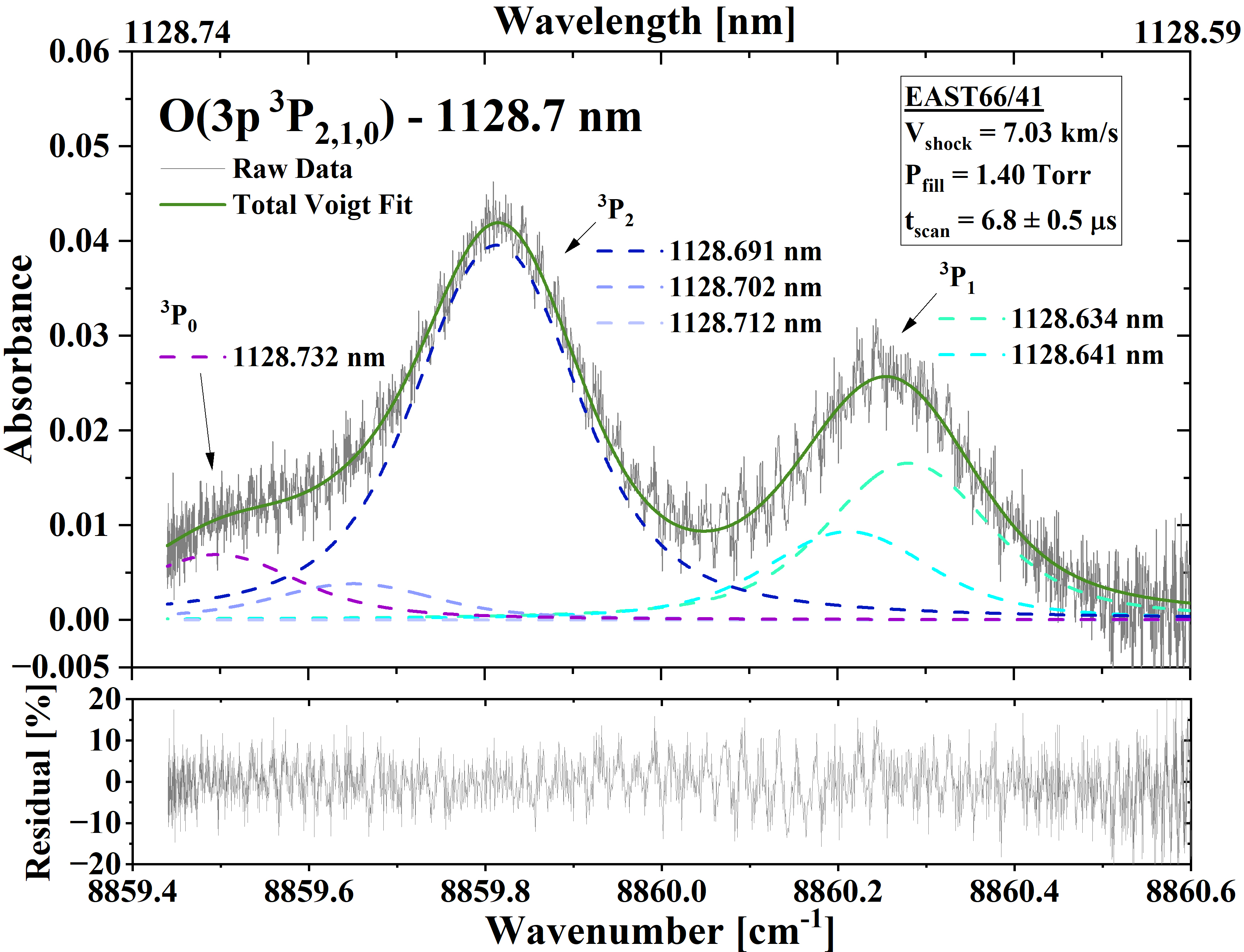}}}%
\caption{\label{fig:excited_fits}Voigt profile fits; (a) O($^5S$) near 777 nm; (b) N($^4P$) near 868 nm; (c)  O($^5P$) near 926 nm; (d) O($^3P$) near 1128 nm. Residuals are presented as a percentage of the maximum absorbance.}
\end{figure}

\begin{figure}[]
\centering
\subfigure[]
{{\includegraphics[width=.49\textwidth]{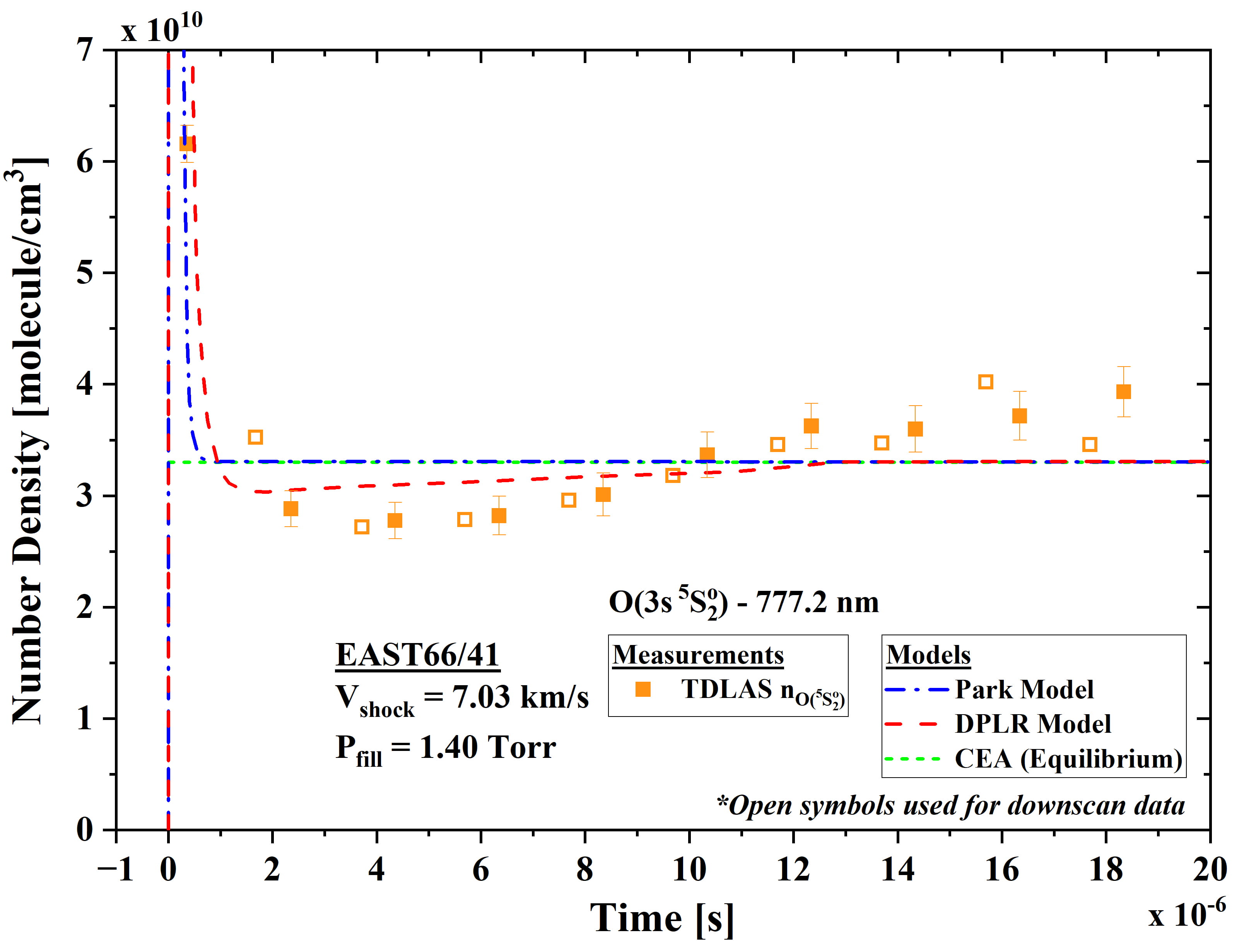}}\label{sfig:777Dens}}%
\subfigure[]
{{\includegraphics[width=.49\textwidth]{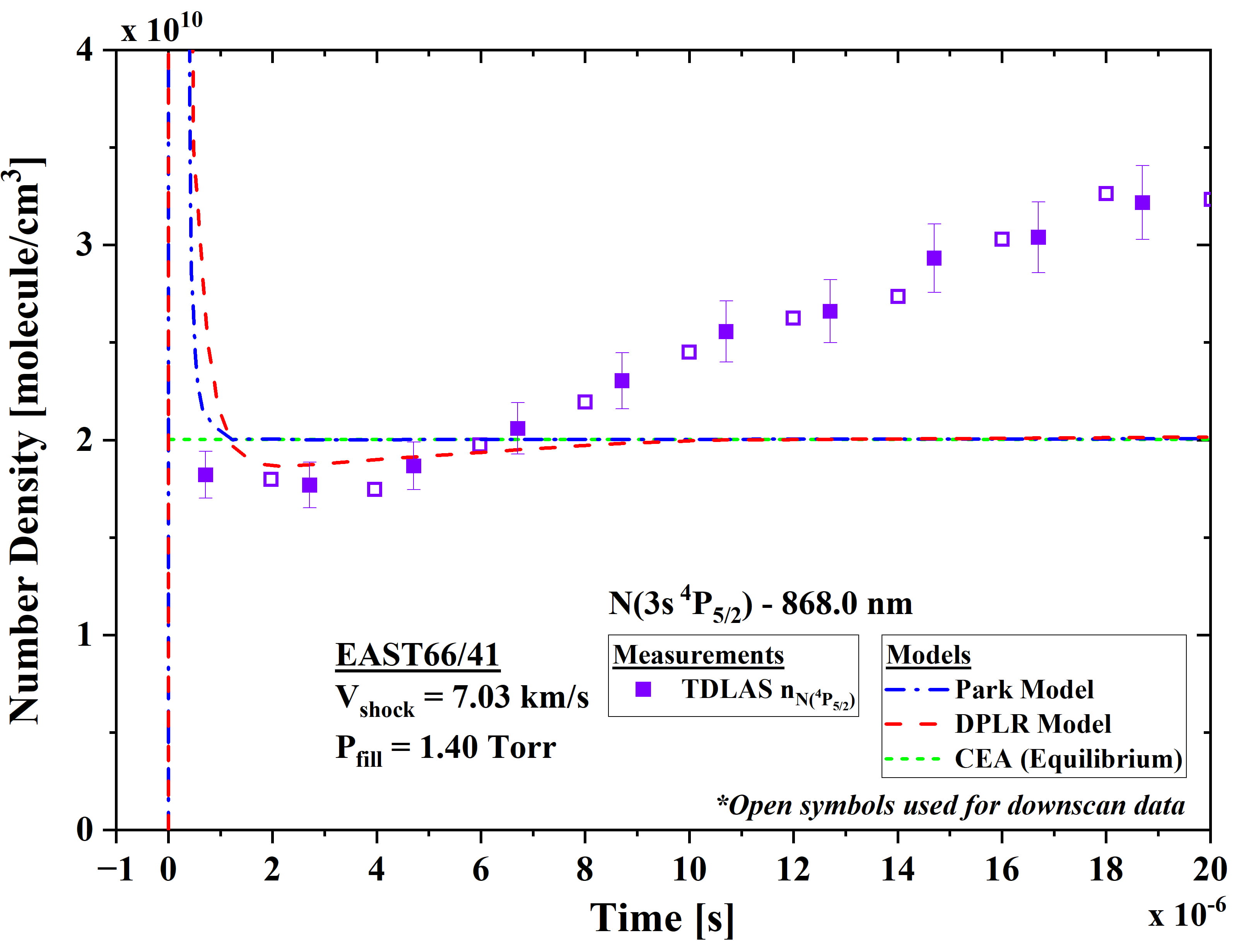}}}\\% 
\vspace{-3mm}
\subfigure[]
{{\includegraphics[width=.49\textwidth]{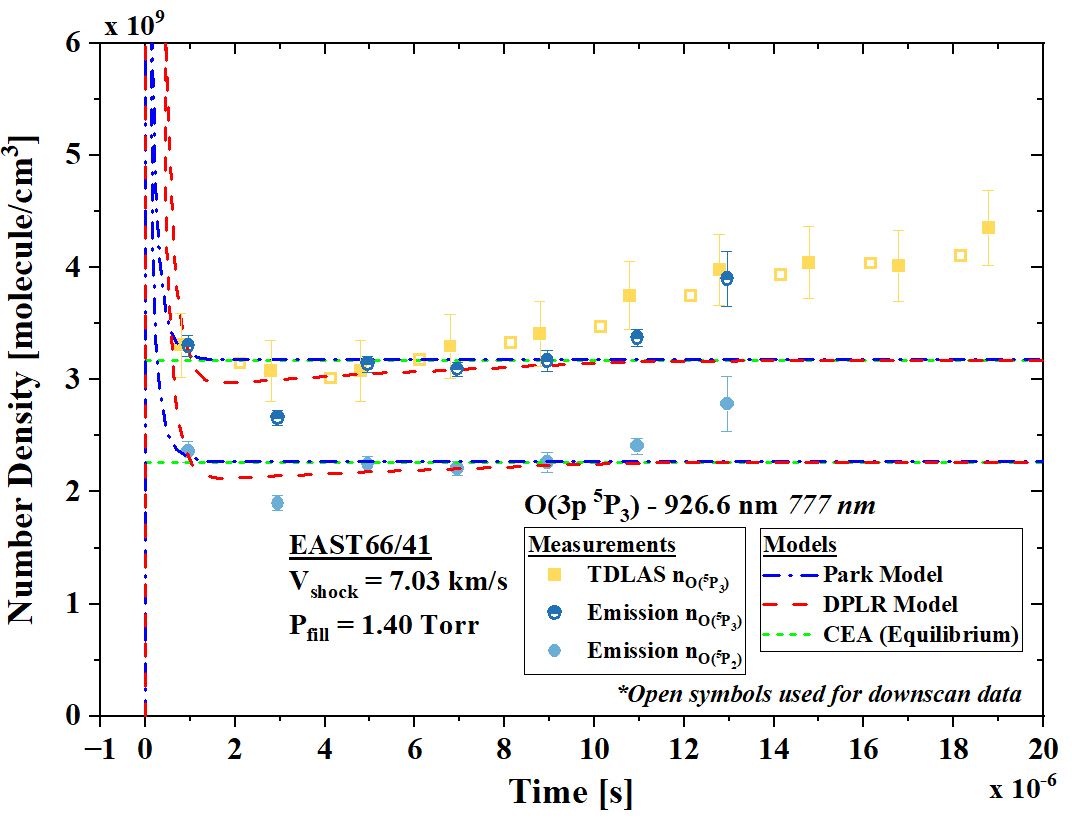}}\label{sfig:926Dens}}%
\subfigure[]
{{\includegraphics[width=.49\textwidth]{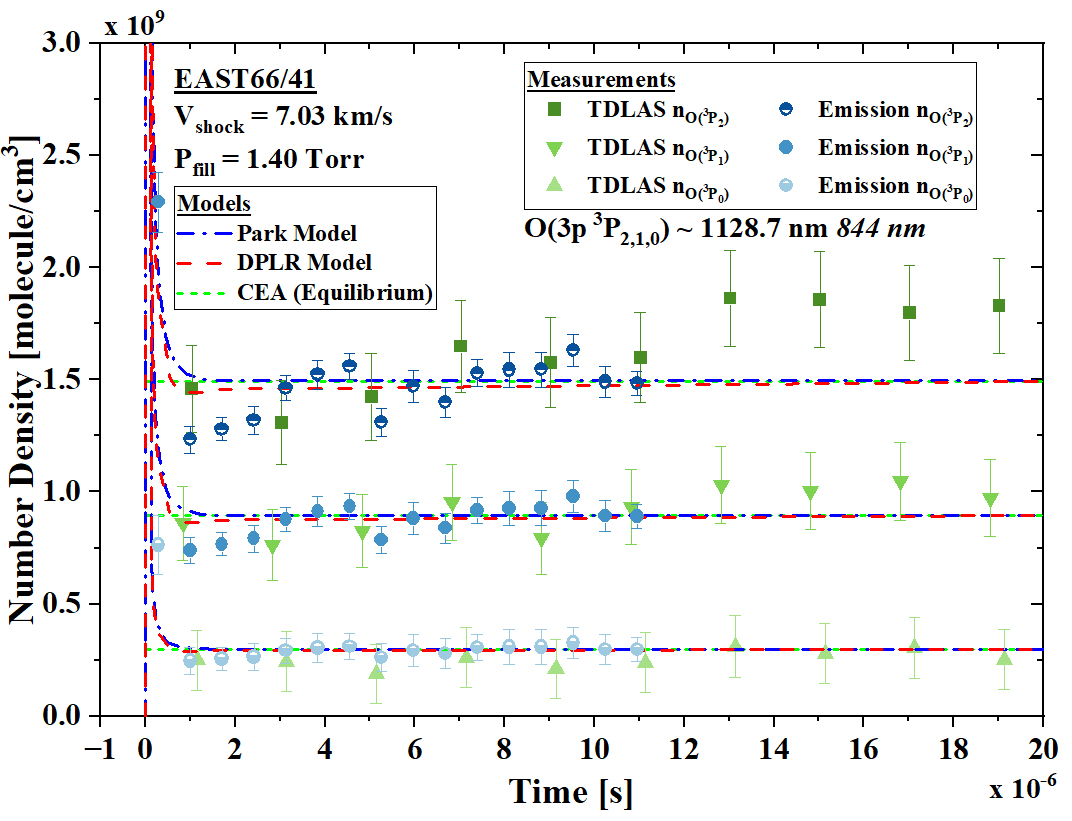}}\label{sfig:1128Dens}}\\%
\vspace{-3mm}
\subfigure[]
{{\includegraphics[width=.49\textwidth]{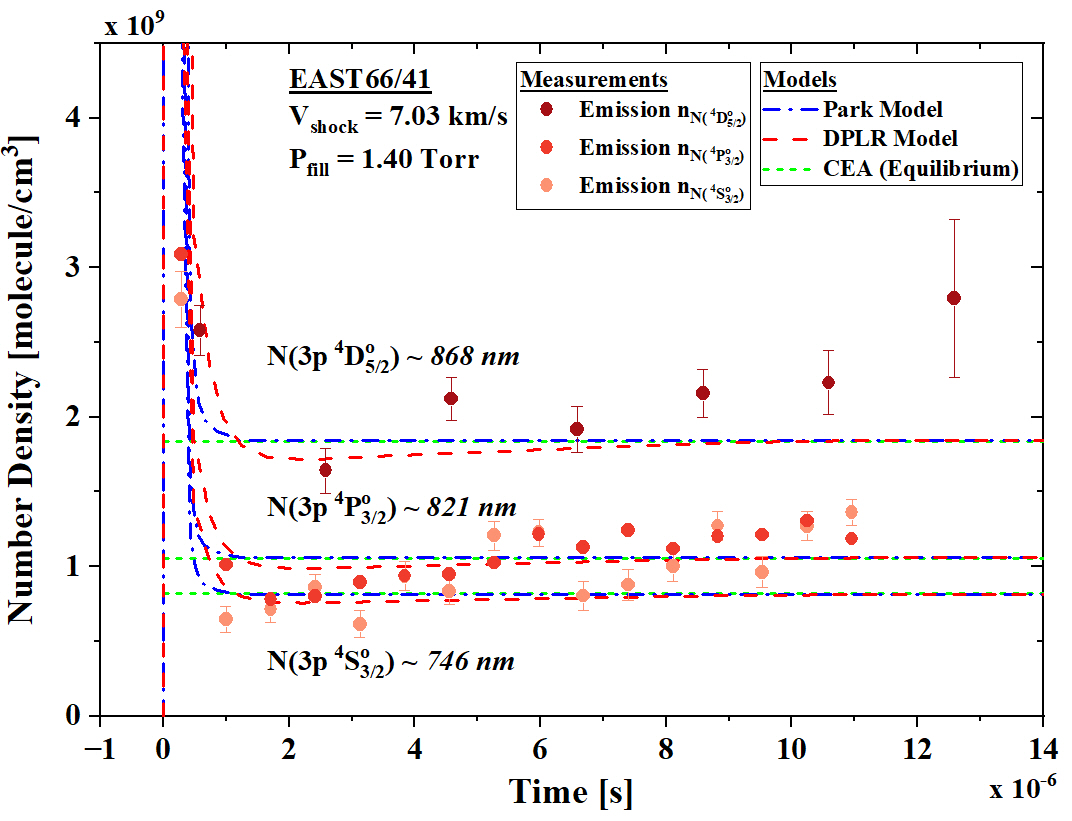}}\label{sfig:NEmissDens}}%
\subfigure[]
{{\includegraphics[width=.49\textwidth]{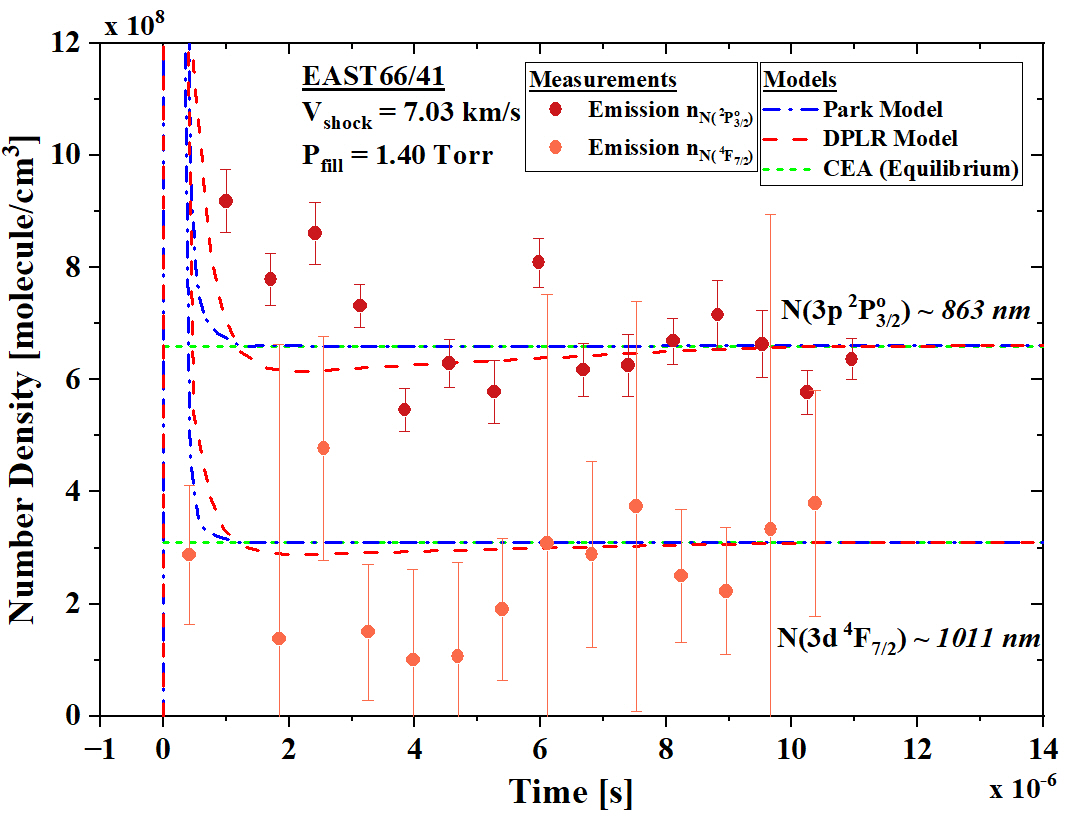}}}%
\vspace{-3mm}
\caption{\label{fig:41_density} Excited state number densities; (a) O($\mathrm{^5S}$); (b) N($\mathrm{^4P}$); (c) O($\mathrm{^5P}$); (d) O($\mathrm{^3P}$); (e) N($\mathrm{^4D}$), N($\mathrm{^4P}$), and N($\mathrm{^4S}$), respectively; and (f) N($\mathrm{^2P}$), and N($\mathrm{^4F}$). Absorption (upright) and emission (italics) wavelengths are listed.}
\end{figure}

\begin{figure}[htb!]
\centering
\subfigure[]
{{\includegraphics[width=.49\textwidth]{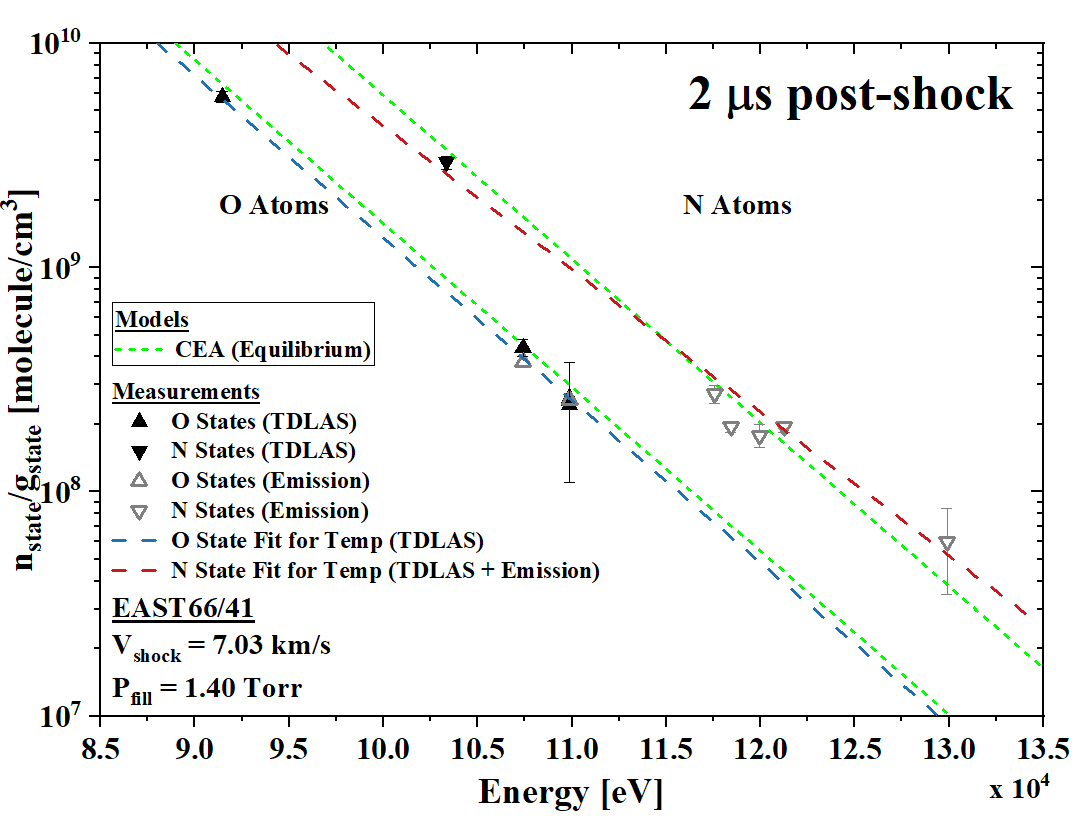}}}%
\subfigure[]
{{\includegraphics[width=.49\textwidth]{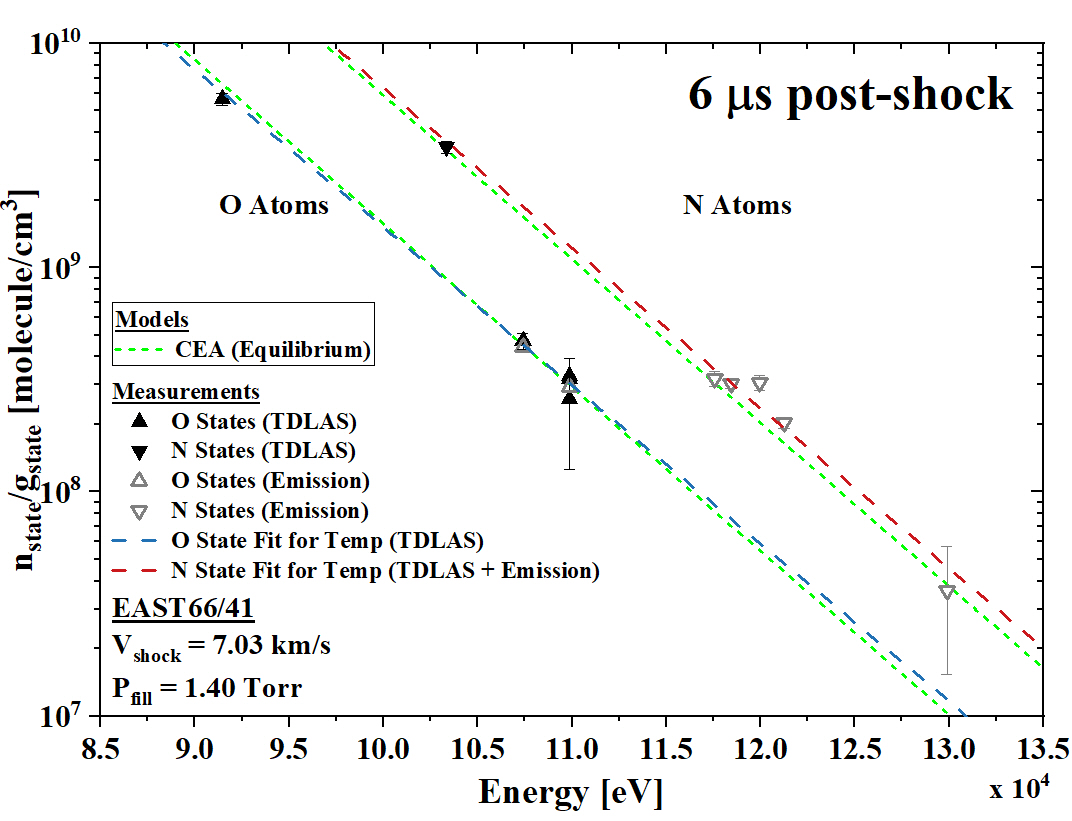}}}\\
\subfigure[]
{{\includegraphics[width=.49\textwidth]{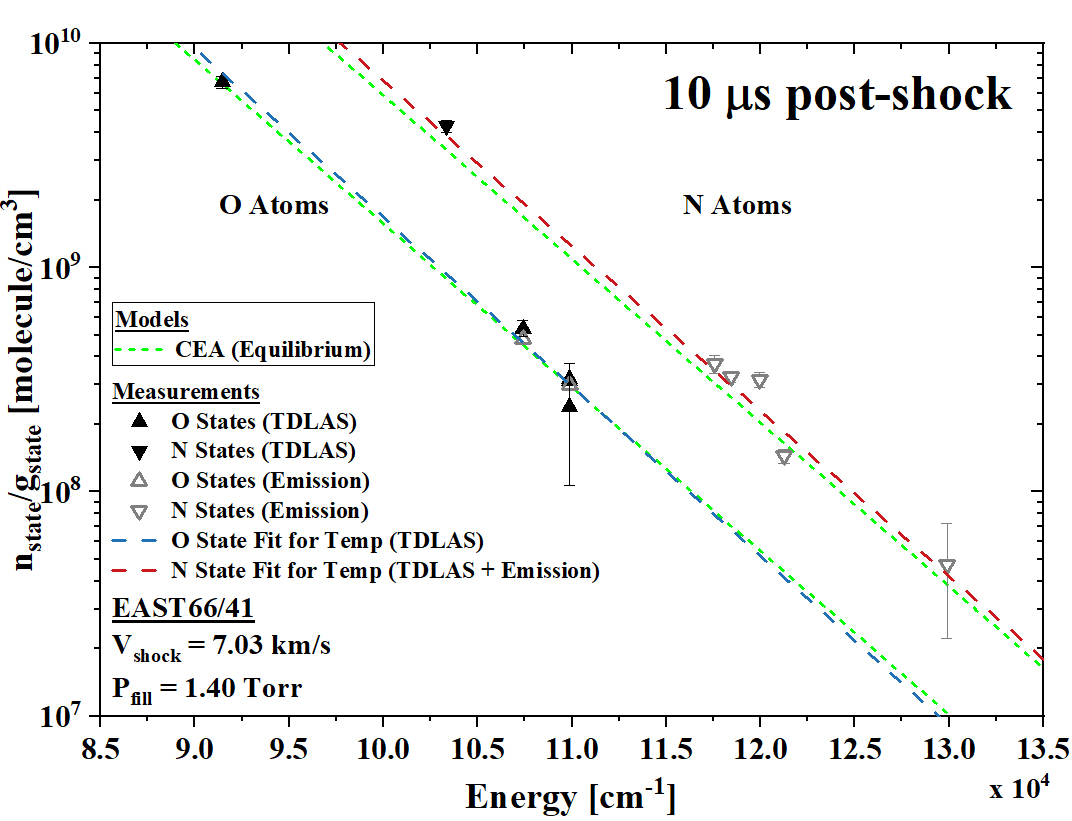}}}%
\caption{\label{fig:boltz_fits}Boltzmann plots of the nine measured atomic levels; (a) 2 $\upmu$s, (b) 6 $\upmu$s, and (c) 10 $\upmu$s post-shock. O and N population temperatures were inferred from the slopes, respectively.}
\end{figure}

\begin{figure}[]
\centering
\subfigure[]
{\includegraphics[width=0.49\textwidth]{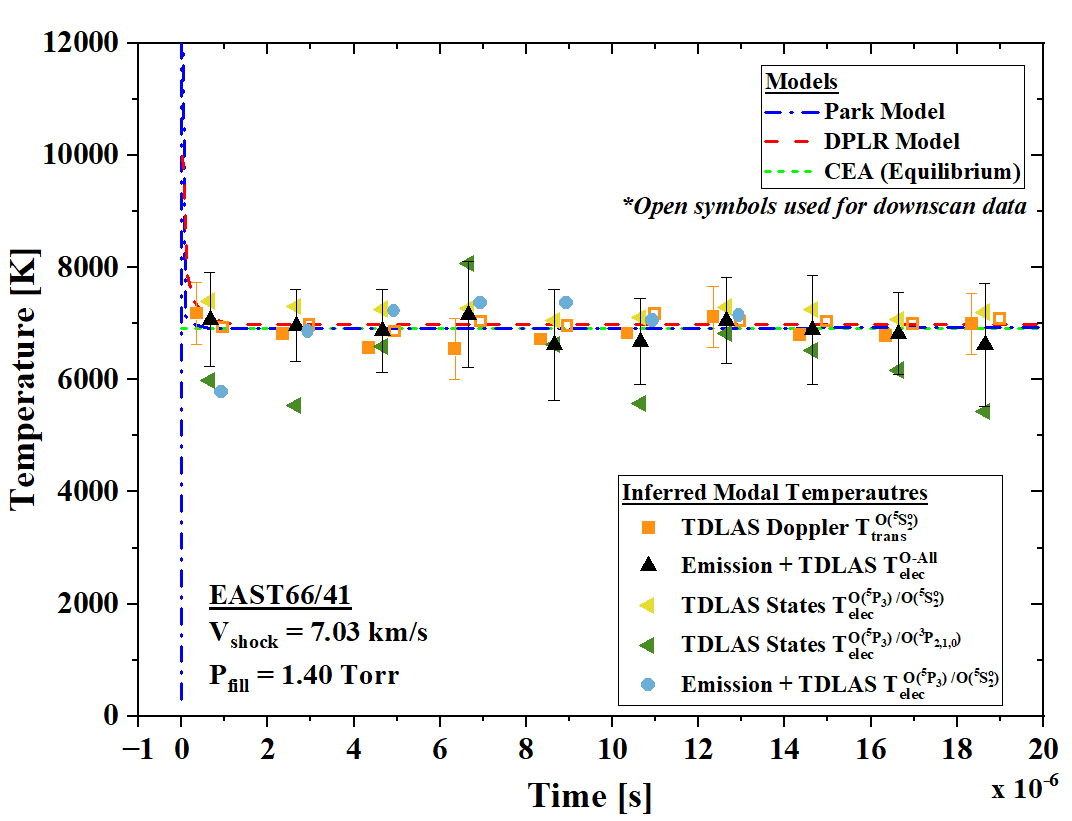} \label{fig:41_TDLAS_temp}}
\subfigure[]
{\includegraphics[width=.49\textwidth]{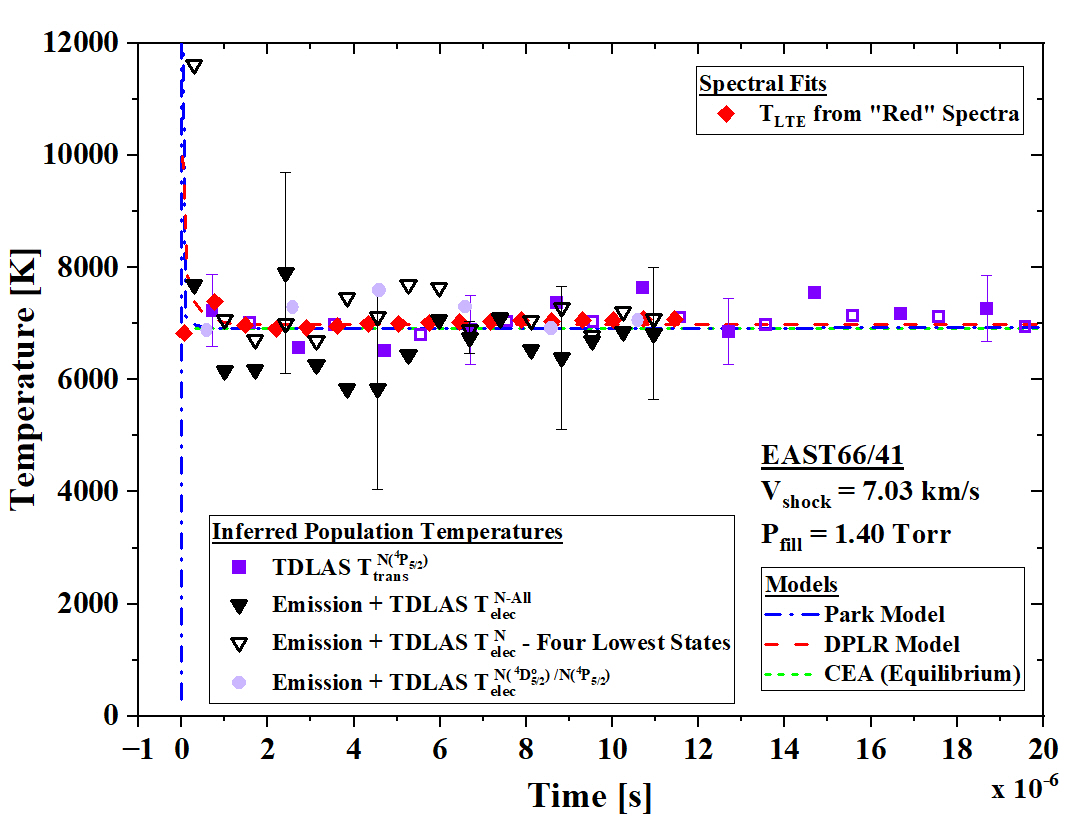}\label{fig:41_OES_temp}}
\caption{\label{fig:41_temp} Shot 66/41 (7.03~km/s, 1.40~Torr): (a) Doppler and electronic temperature inferences from O-atoms; (b) Doppler and electronic temperature inferences from N-atoms as well as from the OES spectral fit.}
\end{figure}

One high-velocity condition is used to demonstrate the resolution of many electronic state populations of O and N in an experiment at 7.03 km/s and 1.40 Torr fill pressure (shot 66/41). This case was part of the second series of shocks at higher velocities (5.9-7.1 km/s) and lower fill pressures (1.4-2.0 Torr) to replicate air data obtained in the Oxford T6 facility for cross-facility validation \cite{oxford, cruden2024b}.
Fig. \ref{fig:excited_fits} shows example absorbance profiles for the four different laser wavelengths -- 777.2 nm, 926.6 nm, and 1128.7 nm for O, and 868.0 nm for N.
As with NO, these absorbance profiles were calculated from the I$_t$ and I$_0$ profiles shown in Fig. \ref{fig:raw} using Eqn. \ref{eq:beer}.
The resulting profiles are similarly fit with Voigt profiles, with 777.2 nm and 868.0 nm exhibiting isolated, single transitions and 926.6 nm and 1128.7 nm being fit with overlapping muliplet transitions using relative linecenter positions from NIST ASD \cite{nist}.
These Voigt fits return both a Doppler broadening width, for $T_{\mathrm{trans}}$ inferences via Eqn. \ref{eq:doppler}, as well as integrated absorbance area for electronic state populations via Eqn. \ref{eq:intA}.
Multiplets were analyzed using the summed integrated area, where the three transitions near 926.6 nm were summed and the six transitions near 1128.7 nm grouped into the three closely spaced states (e.g. $ \specnot{3p}{3}{P}{2,1,0}{} $).
Thus, TDLAS data return both state populations (Fig. \ref{fig:41_density}) and $T_{\mathrm{trans}}$ (Fig. \ref{fig:41_TDLAS_temp}), with the distribution of state populations enabling inference of $T_{\mathrm{elec}}$ (Fig. \ref{fig:boltz_fits}).

\begin{figure}[]
	\centering
    \subfigure[]
{{\includegraphics[width=.49\textwidth]{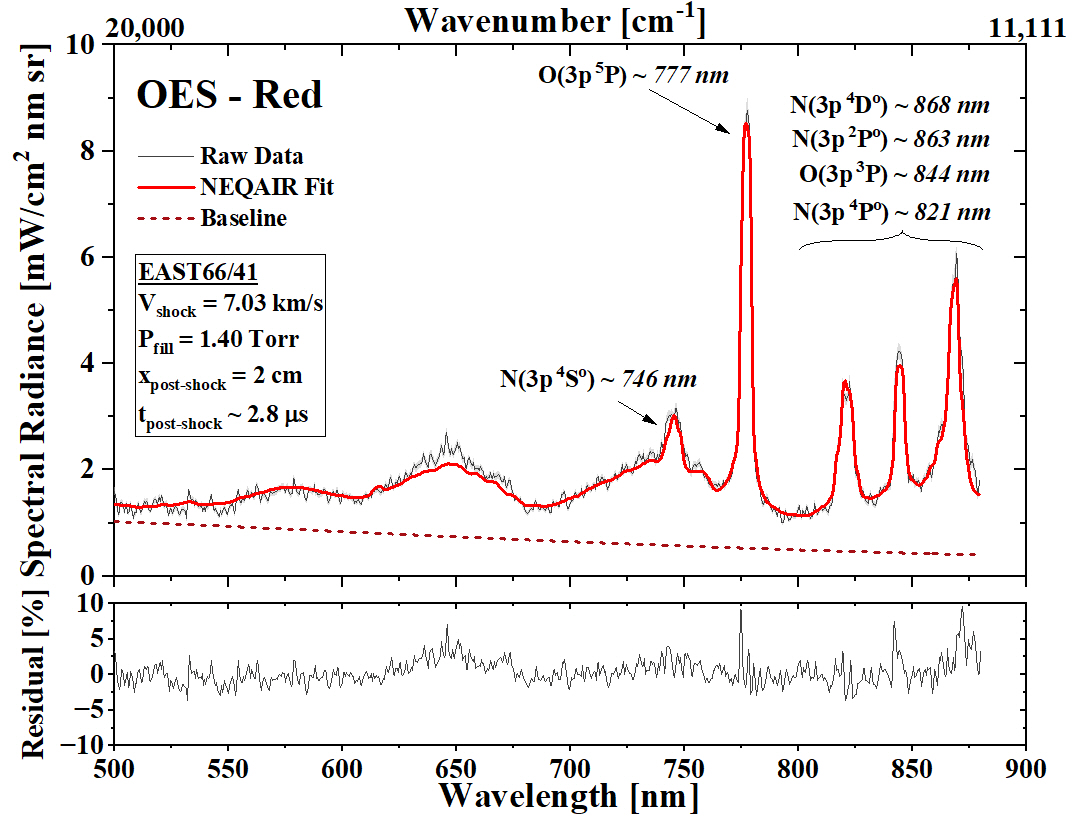}}}%
	\subfigure[]
{{\includegraphics[width=.49\textwidth]{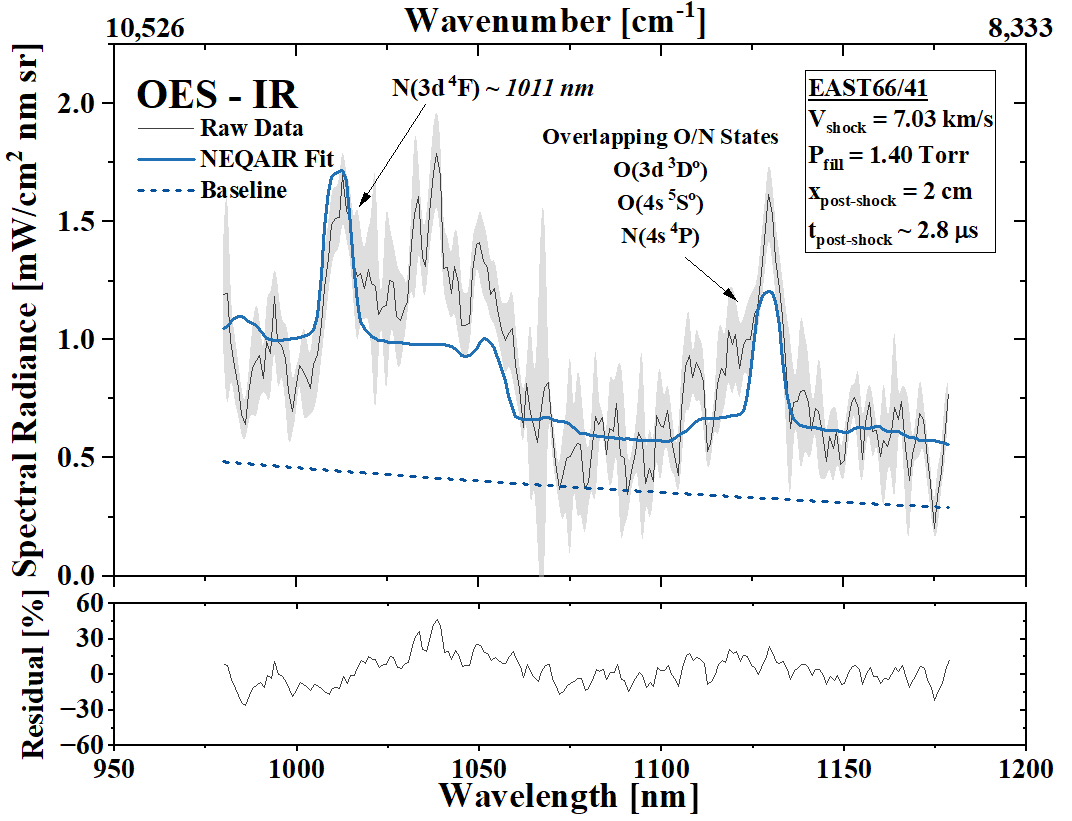}}}%
			
	\caption{Comparison between measured and fit OES spectra for shot 66/41 at 2~cm after the shock for the (a) Red and (b) IR cameras. Residuals are calculated as a percentage of the maximum NEQAIR-predicted radiance.}
	\label{fig:EAST66-41_fits_LTE}
\end{figure}

The corresponding OES measurements from the Red and NIR cameras for shot 66/41 at $\SI{2.8}{\micro\second}$ after shock arrival are also presented in Fig.~\ref{fig:EAST66-41_fits_LTE}. 
Spectral fits were performed using NEQAIR v15.3, assuming equilibrium composition at the fit temperature and the post-shock equilibrium pressure; these results are also shown in the figure.
A Planck-like spectrum, expressed as $\chi_{\ce{BB}} L_{\ce{BB}}(T)$ and evaluated at the fitted temperature, was used as a convenient representation of the background continuum, as shown by the dashed lines. The scaling parameter $\chi_{\ce{BB}}$ was also determined as part of the fitting procedure.
Such a continuum contribution is commonly observed in shock-tube OES measurements \cite{cruden2009, glenn2024}, and leads to an offset between predicted and measured spectra.
Equilibrium spectral fits of the \redcam\ camera were also conducted up to 10~cm after the shock location at 0.25~cm intervals, with 0.25~cm spatial bins, and reported in Fig.~\ref{fig:41_OES_temp} (e.g. data from 0 - $\SI{12}{\micro\second}$). 
The equilibrium fits show good agreement with both the \redcam\ and \ircam\ cameras, with residuals largely within the estimated uncertainty bounds, confirming that the shocked gas has effectively reached equilibrium.
OES spectra provide two methods for inferring temperature.
First, the "Red" and "IR" camera data can return an overall temperature fit from NEQAIR, with data in Fig.~\ref{fig:EAST66-41_fits_LTE} specifically corresponding to Red: $T\subrm{LTE} = \SI{6942}{\kelvin}$, $\chi_{\ce{BB}} = 1.7\times 10^{-4}$ and IR: $T\subrm{LTE} = \SI{7015}{\kelvin}$, $\chi_{\ce{BB}} = 2.6\times 10^{-4}$, where $T\subrm{LTE}$ denotes local thermodynamic equilibrium.
Thus, the full spectral fits are used to infer an overall temperature fit, termed "$T_{\mathrm{LTE}}$ " in Fig. \ref{fig:41_OES_temp}, with the "Red" camera fits retained, as these spectra contain the majority of the relevant OES features.
The data analysis for radiance from each emission feature (Eqns. \ref{eq:OES_nu_over_gu}-\ref{BoltzTemp}) was also used for additional state populations and $T_{\mathrm{elec}}$ inferences (Figs. \ref{fig:41_density}-\ref{fig:41_temp}).
In this case, a linear baseline subtraction was performed around each line in order to offset both the continuum and molecular backgrounds. 

Using the "Red" camera, populations of the  O($ \specnot{3p}{5}{P}{}{} $) and O($ \specnot{3p}{3}{P}{}{} $) states were measured from transitions around 777~nm and 844~nm, respectively, while populations of the N($ \specnot{3p}{4}{S}{}{\circ} $), N($ \specnot{3p}{2}{P}{}{\circ} $), N($ \specnot{3p}{4}{P}{}{\circ} $), and N($ \specnot{3p}{4}{D}{}{\circ} $) states were measured from transitions around 746~nm, 821~nm, 863~nm, and 868~nm respectively.
The 863~nm and 868~nm features overlap, with the strength of 868~nm emission substantially exceeding that of 863~nm.
However, data processing could reasonably separate the two contributions, albeit with fairly large noise for the N($ \specnot{3p}{4}{P}{}{\circ} $) state at 863~nm due to the nearby strong 868~nm emission.
From the "IR" camera, an additional N-atom state -- N($ \specnot{3d}{4}{F}{}{} $) was measured from the transition around 1011~nm.
The feature present at 1128~nm includes contributions from three overlapping O and N states, which precluded inferences of any of the three states, since the transition could not be reliably isolated from neighboring O and N transitions with the available OES resolution.
However, the NEQAIR v15.3 simulation  adequately captures this combined feature.

Ultimately, the OES state number density inferences are affected by the spatial resolution of the OES system, as well as by the spatial binning of $\SI{0.25}{\cm}$ employed when fitting the spectra, thus representing a value corresponding to the average radiance.
While TDLAS can resolve closely spaced multiplet states, OES data do not have the spectral resolution to separate closely spaced features -- instead observing combined features.
For each isolated emission feature, inferences of state populations from OES data were not influenced by overlapping multiplet states, as they were treated by assuming n$_j$/g$_j$ as approximately constant.
Ultimately, the OES spectra demonstrate the simultaneous resolution of emission features arising for a variety of O- and N-atom electronic states, albeit with coarser resolution of any individual feature than the TDLAS data.

TDLAS and OES greatly complement each other, as TDLAS can help address optical thickness assumptions used for OES data processing, while OES probes many more states in a given experiment.
The optical thickness of the probed emission lines was simulated with NEQAIR v15.3, indicating a relevant impact for the O($ \specnot{3p}{5}{P}{3}{} $) and N($ \specnot{3p}{4}{D}{}{0} $) transitions at the observed resolution. However, these two states are upper states of the 777~nm and 868~nm transitions respectively, so their lower state populations could be leveraged from the TDLAS data. This further motivated the use of Eq. \ref{eq:OES+TDLAS}, which accounts for this effect by the use of combined OES--TDLAS measurements. For transitions lacking TDLAS measurements, this optical thickness was typically found to be negligible, and Eq.~\ref{eq:OES_nu_over_gu} was therefore applied. The O 844~nm lines constitute an exception, as a non-negligible effect is expected but could not be corrected owing to the absence of corresponding TDLAS data.
Together, states probed by OES supplement those probed by TDLAS resulting in three O-atom state and six N-atom state measurements, with two states -- O($ \specnot{3p}{3}{P}{2,1,0}{} $) and O($ \specnot{3p}{5}{P}{3}{} $) -- probed via both emission and absorption (summarized in Fig. \ref{fig:energylvls}).
Thus, out of the nine probed electronic states, two states are probed with both absorption and emission, providing another direct comparison of the techniques.

The combined TDLAS and OES electronic state population number density data are compiled in Fig. \ref{fig:41_density}.
Each of the TDLAS states is plotted separately, with Figs. \ref{sfig:926Dens} and \ref{sfig:1128Dens} showing the overlapping TDLAS and OES probing of the O($ \specnot{3p}{3}{P}{2,1,0}{} $) and O($ \specnot{3p}{5}{P}{3}{} $) states.
The five OES N-atom states grouped into two plots, where N($ \specnot{3p}{4}{S}{}{\circ} $), N($ \specnot{3p}{2}{P}{}{\circ} $), and N($ \specnot{3p}{4}{D}{}{\circ} $) are grouped due to their shared lower state -- i.e. the N($ \specnot{3s}{4}{P}{5/2}{} $) state probed with TDLAS at 868~nm.
TDLAS data use the same color scheme as in Fig. \ref{fig:excited_fits}, while the OES data are colored blue for O-atom states and red for N-atom states.
Furthermore, the TDLAS data use filled symbols for upscan data and open symbols for downscan data -- as previously described.
In contrast the OES data are plotted with filled symbols for the average degeneracy (i.e. the middle value of g$_j$ for multiplet states), as these values reflect the output of the OES data analysis.
However, these values are scaled by the degeneracy ratios to compare with TDLAS data at other degeneracies (e.g. n$_{O(^5P_3)}$/n$_{O(^5P_2)}$ = 7/5).
This scaling is supported by the observations of O($ \specnot{3p}{3}{P}{2,1,0}{} $), where the three degeneracies measured with TDLAS exhibit the expected proportionality (Fig. \ref{sfig:1128Dens}).
Finally, two colors of lines are used for the Park and DPLR models, with the CEA-calculated equilibrium value plotted again in light green. The Streicher \textit{et al.} model does not include ionization and so is not used for comparison for the O-atom and N-atom states.

All the measured states form very quickly, and the resulting values remain close to the predicted post-shock equilibrium from about $\SI{2}{\micro\second}$ from the shock front onward, with the exception of the earliest time data points in Figs. \ref{sfig:777Dens}, \ref{sfig:926Dens}, and \ref{sfig:NEmissDens}.
In these cases, there is some evidence of higher O and N state density due to post-shock excitation and ionization chemistry, although the time resolution is insufficient to adequately capture the rapidly changing number density.
However, all data exhibit a slowly increasing number density at late times, consistent for all states and for both TDLAS and OES data.
This increasing trend was consistent for both techniques and was observed at other high-speed conditions in Test Series 66, likely caused by shock trajectory and boundary layer compression effects.
Regardless, this increase in atomic state number densities is consistent across all states, with the ensemble of states maintaining approximately the same distribution over this time.

To demonstrate this evolution of state populations and depict the method for extracting $T_{\mathrm{elec}}$, these number density results are plotted in a Boltzmann plot for post-shock data at approximately $\SI{2}{\micro\second}$, $\SI{6}{\micro\second}$, and $\SI{10}{\micro\second}$ in Fig. \ref{fig:boltz_fits}.
For each time instance, O-atom and N-atom states were separately fit subject to the Boltzmann distribution description in Eq. \ref{BoltzTemp} to infer $T_{\mathrm{elec}}$ for each species, which are represented with the red and blue dashed lines.
For comparison, the distribution for the CEA equilibrium O- and N-atom number density and equilibrium temperature is repeated in each plot as a light green line.
This CEA value is unchanged between the three plots, and helps to discern the overall rise in O- and N-atom number densities, which is slightly obscured due to the logarithmic scale.
Thus, Fig. \ref{fig:boltz_fits} provides a visualization of the state population distributions and their evolution over time.
Notably, the O-atom states remain distributed with a slope quite close to the expected CEA temperature, while the N-atom states have more variable slope over time.
This variation -- and hence scatter in the inferred N-atom $T_{\mathrm{elec}}$ is strongly influenced by the two highest-lying states, which exhibit larger scatter in their state populations than the four lower-lying states.
The behavior of high-lying states could be more strongly influenced by collisional-radiative processes, although this would require more substantial modeling \cite{Aiken_CollRad,Aiken_2025,Bourdon_1998,Jo_2019}.
However, when N-atom $T_{\mathrm{elec}}$ inferences are restricted to just the four lowest states, the temperature inferences remain in much better agreement with the CEA equilibrium value (Fig. \ref{fig:41_temp}).

Measurements of nine electronic states, along with the temperature dependent spectral data, provide a wide variety of temperature inferences.
Thus, temperature time-histories for shot 66/41 (7.03 km/s, 1.4 Torr) are provided in Fig.~\ref{fig:41_temp} for many different temperature interpretations.
Data in Fig. \ref{fig:41_TDLAS_temp} include temperature inferences for O-atoms that leverage the TDLAS data primarily  - including the best-fit Doppler temperature ($T_{\mathrm{trans}}$) for the well-isolated 777.2 nm feature as well as O-atom $T_{\mathrm{elec}}$ for various combinations of the measured electronic states.
Doppler temperatures include downscan data (open symbols), while electronic temperatures were only available for upscan data due to the scan characteristics of the 1128.7 nm laser.
The larger scatter and uncertainty in the O($\mathrm{^5P_3}$)/O($\mathrm{^3P_{2,1,0}}$) inference of $T_{\mathrm{elec}}$ (green) is expected due to the relatively smaller energy spacing of these higher-lying excited states (Fig. \ref{fig:boltz_fits}).
The addition of emission measurements at 777 nm was also used to infer $T_{\mathrm{elec}}$ values (light blue), which differs slightly from the TDLAS only inference (yellow) due to slight early-time discrepancies of O($ \specnot{3p}{5}{P}{3}{} $) inferences between the two methods.
Finally, the black data points show the O-atom $T_{\mathrm{elec}}$ inferred from all states (Fig. \ref{fig:boltz_fits}).
Together, all O-atom temperature data remain relatively self-consistent, and all inferences tend to reflect the expected CEA temperature, even if the O-atom number densities slightly exceed the CEA prediction.

Data in Fig. \ref{fig:41_OES_temp} include N-atom temperature inferences from a broader combination of OES and TDLAS data.
The value of combining OES and TDLAS is immediately apparent from the additional N-atom $T_{\mathrm{elec}}$ inferences that were unavailable from TDLAS data alone.
In fact, the only temperature inference available from TDLAS for N-atoms is the Doppler temperature ($T_{\mathrm{trans}}$) for the well-isolated 868.0 nm feature.
Fig. \ref{fig:41_OES_temp} also includes the $T_{\mathrm{LTE}}$ values inferred from the spectral fit of "Red" camera data (red data points) that arises solely from OES data.
These $T_{\mathrm{LTE}}$ values fit across both O-atom and N-atom features, but are included in \ref{fig:41_OES_temp} for comparison.
Furthermore, three electronic temperature results are provided for N-atom states.
First, fits over all six of the electronic states are given via filled black symbols.
For these fits, the two highest-lying N-atom states exert a strong influence on reducing $T_{\mathrm{elec}}$ inferences to be below the CEA equilibrium value.
This is apparent when comparing with the $T_{\mathrm{elec}}$ fit from just the four lowest-lying states, with results in much better agreement with the CEA value.
Finally, the two states involved in the 868 nm transition (N($ \specnot{3p}{4}{D}{5/2}{\circ} $) and N($ \specnot{3s}{4}{P}{5/2}{} $)) were used to infer a specific $T_{\mathrm{elec}}$ for the two states, with this value in relatively good agreement with CEA as well.
The many resolved electronic states thereby yield many possible inferences of temperature from the TDLAS and OES data that explore the many possible collisional and radiative interactions among these electronic states.

As is, the models do not directly output excited state populations.
Rather, state-specific model predictions in Fig. \ref{fig:41_density} are inferred from the output total O- and N-atom populations and an assumed Boltzmann distribution at the translational temperature.
Furthermore, the timing of the nonequilibrium region is such that all model-predicted nonequilibrium processes are complete well within the first $\SI{1}{\micro\second}$.
As was noted, a few of the earliest TDLAS and OES data may observe part of this nonequilibrium, but the time resolution remains a challenge for these very high-speed shock experiments.
Future work could investigate the lower velocity shocks (i.e. 5.9 km/s) where nonequilibrium lasts longer.
The ensemble of 5.9-7.1 km/s data substantially extend the application of TDLAS and OES in air experiments to include excitation and ionization processes, with the combined diagnostic strategy again demonstrating opportunities to investigate nonequilibrium and facility effects in EAST experiments.

%=======================================================================================================
\section{Conclusions}

Two TDLAS-based sensor systems targeting NO in the MIR and excited states of O and N in the NIR were employed in the post-incident-shock region at the NASA EAST facility for a sweep of conditions representative of Earth reentry. For the low-velocity (<5 km/s) equilibrium condition (shot 66/21), the TDLAS NO diagnostic yields modal temperatures and NO number densities that agree with models within uncertainty. At the non-equilibrium condition (shot 66/26), TDLAS measurements trend well with models, falling into closest agreement with the Streicher \textit{et al.} model. Discrepancy between model predictions are most prominent in the early times, and may be traced to differing reaction rates (O$_2$ dissociation, Zeldovich). More detailed investigation is reserved for future work. In the high-velocity (>5 km/s) condition (shot 66/41), TDLAS-inferred Doppler temperatures collapse to equilibrium within the first microsecond of test time, in agreement with models. Furthermore, the four NIR laser wavelengths probe three O-atom states and one N-atom state for comparison with both OES data and kinetic models. This work demonstrates the successful application of TDLAS systems for air experiments at EAST over a wide range of velocity conditions, demonstrated through the three cases presented here, for inferring time-resolved modal temperature and species number density profiles.

The combination of TDLAS and OES data substantially extends the diagnostic capability at EAST, while enabling a comparison between the two techniques. OES data provide a direct comparison with upper states of a transition, which pairs well with the ground-state TDLAS measurements. For NO, this produced similar NO number density and modal temperatures as were available from TDLAS, as both IR measurements probe rovibrational states of the ground electronic state of NO. The resulting NO temperature and number density time-histories exhibit reasonable agreement between TDLAS and OES, although TDLAS tends to exhibit lower scatter and uncertainty -- as can be expected from the reduced emission signal for low-speed shocks. The trends for both techniques capture the temperature decline fairly well, and both show a higher production of NO than predicted by the baseline Park93 model. Disagreement between the techniques is largest for NO number densities in the equilibrium condition (shot 66/21), although this can be explained by spatial averaging in the OES data that is not present for TDLAS. For O- and N-atoms, the OES data greatly extend the number of available electronic states probed in each experiment, enabling simultaneous measurements of three O-atom states and six N-atom states. The resulting electronic state number densities and corresponding modal temperatures remain close to equilibrium, with some evidence of depleted electronic state populations for the highest-lying N-atom states. Furthermore, all electronic state number densities slowly rise, again suggesting potential facility effects (i.e. boundary layers and shock trajectory) that influence the EAST experimental observations. Thus, the combined diagnostic strategy provides valuable new capabilities for EAST experiments over a very wide range of conditions.

Future work will seek to provide enhanced temporal resolution in the early test-time, nonequilibrium region for the high-speed shocks, likely through a combination of equipment upgrades and more advanced laser electronics (i.e. a bias-tee configuration for TDLAS). Beyond diagnostic improvements, additional Test Series 66 data analysis can be used to investigate nonequilibrium and facility effects. In particular, the longer recording time of TDLAS data can be used to compare with boundary layer models for incident shock facilities, while the full set of velocity conditions can be used to evaluate kinetic model performance over a wide range of conditions. Notably, these additional observations should provide valuable comparisons to similar reflected shock studies of air nonequilibrium \cite{StreicherVibAir_PartI,StreicherVibAir_PartII,StreicherVibAir_PartIII}. Finally, the TDLAS/OES combination could be applied to other atmosphere compositions (e.g. Mars/Venus or Titan) to support NASA missions beyond Earth's atmosphere. Ultimately, this work successfully demonstrated two fast-scanning laser absorption diagnostics in EAST and their combination of this data with the existing EAST OES systems, yielding inferences of temperatures for multiple internal energy modes and quantum-state-specific population number densities through a sensitive probing of NO, O and N, with many potential opportunities to extend measurements in future EAST experiments.

\section*{Funding Sources}
Partial support of this research was provided by AFOSR grant numbers FFA9550-19-1-0219 and FA9550-21-1-0311, with contract monitors Drs. Ivett Leyva, Sarah Popkin, and Michael Berman. E. Chang was supported by a NASA Space Technology Research Fellowship. The research of A. C. Tibere-Inglesse and A. Fagnani was supported by an appointment to the NASA Postdoctoral Program at the NASA Ames Research Center, administered by Oak Ridge Associated Universities under contract with NASA. B. A. Cruden was supported by NASA contract NNA15BB15C to Analytical Mechanical Associates, Inc. The test campaign was supported by the NASA Entry Systems Modeling project.

\bibliography{air_bib}

\end{document}